\documentclass[twocolumn,revtex4,apj,iop, twocolappendix, numberedappendix]{openjournal}
\usepackage{xcolor}
\usepackage{graphicx}
\usepackage{amsfonts}
\usepackage{amssymb, bm}
\usepackage{url}
\usepackage[breaklinks,colorlinks,citecolor=blue,linkcolor=blue,urlcolor=blue]{hyperref}
\usepackage{amsmath}
\usepackage[varg]{txfonts}
\usepackage[normalem]{ulem}
\usepackage{fontawesome}

\DeclareMathAlphabet{\mathbbold}{U}{bbold}{m}{n}

\renewcommand{\d}{\mathrm{d}}
\newcommand{\p}{\partial}

\newcommand{\e}{\mathrm{e}}

\newcommand{\s}{\mathrm{s}}
\renewcommand{\o}{\mathrm{o}}

\setcitestyle{numbers,square}
\begin{document}

\title{Transverse Velocities in Real-Time Cosmology: Position Drift in Relativistic N-Body Simulations}

\author{Alexander Oestreicher$^{\;a,}$\footnote{alexo@cp3.sdu.dk}}
\author{Chris Clarkson$^{\;b,\,c}$}
\author{Julian Adamek$^{\;d}$}
\author{Sofie Marie Koksbang$^{\;a}$ \vspace{0.15cm}}

\affiliation{${}^a$ CP3-Origins, University of Southern Denmark, Campusvej 55, DK-5230 Odense M, Denmark}
\affiliation{${}^b$ Department of Physics \& Astronomy, Queen Mary University of London,  Mile End Road, London E1 4NS, UK\vspace{0.05cm}}
\affiliation{${}^c$ Department of Physics \& Astronomy, University of the Western Cape, Cape Town 7535, South Africa\vspace{0.05cm}}
\affiliation{${}^d$ Institut f\"ur Astrophysik, Universit\"at Zürich, Winterthurerstrasse 190, 8057 Z\"urich, Switzerland \vspace{0.15cm}}

\begin{abstract}
The era of real-time cosmology has begun. It is now possible to directly measure the apparent drift of high-redshift astronomical sources across the sky {\em in real time}. This so-called {\em position drift} provides a valuable probe of the peculiar velocity field and cosmic structure formation by giving direct access to the transverse velocity, which is notoriously difficult to measure and is typically inferred statistically from the density field in a model-dependent way. To fully exploit this new window into the Universe, it is essential to understand how cosmological structures affect position drift measurements. Here we present the first position drift study based on the general relativistic N-body simulation code \texttt{gevolution}. We calculate the position drift directly from the past light cone for ten different observers and compare the results to predictions from linear perturbation theory. At linear order, the position drift is directly proportional to the transverse velocity on the sky. This linear approximation reproduces our non-linear simulation results to within about 5\%. We calculate power spectra for the position drift, splitting the signal into an E- and B-mode and compare the former to linear expectations, finding good agreement. The B-mode is suppressed on linear scales, but has similar amplitude as the E-mode on non-linear scales. We further demonstrate that light-cone inhomogeneities induce biases in the dipole of the drift, introducing redshift dependence of both the amplitude and direction. Although our analysis is not yet sufficient for a firm conclusion, our results suggest that these effects alone cannot explain the possible redshift-dependent dipole in Gaia DR3 data reported in the literature. 
\end{abstract}

\maketitle % OJA
\tableofcontents

\section{Introduction}
Mapping the large-scale peculiar velocity field is a key objective in cosmology. Peculiar velocities respond directly to the underlying gravitational field, making them a uniquely sensitive probe of cosmology, capable of testing gravity on large scales, constraining the growth rate of structure, and probing the nature of dark matter and dark energy \cite{Turner2024}. Despite this importance, measuring peculiar velocities is notoriously difficult. Traditional methods, such as those used in the compilation catalogue CosmicFlows \cite{Tully2023_CF4} or in the DESI Peculiar Velocity Survey \cite{Said2025_Desi_PecVel}, require estimates of distances to the sources. These distances generally have large uncertainties and often rely on model assumptions, such as adopting a $\Lambda$CDM cosmology as the background. Furthermore, the actual extraction of peculiar velocity fields from this type of data requires heavy modelling (see, e.g. \cite{Courtois2023} and references therein).
\newline\indent
A promising complementary approach to the traditional method is to use the real-time observable known as position drift \cite{Quercellini2012_Real_Time_Cosmology, Nusser2012}. Real-time cosmology has the advantage that it directly probes the geometry and kinematics of the underlying spacetime without relying on model assumptions.
\newline\indent
Position drift measures the observed change in the angular position of distant objects over time. In an FLRW universe, it is zero at the background level, but contains information on peculiar velocities, the gravitational wave background \cite{Gwinn1997, Darling2018a}, and potential large-scale anisotropy \cite{Quercellini2009_Cosmological_Parallax, Marcori2018}. On top of the cosmological signal there is also a local contribution from our motion with respect to the cosmic rest frame \cite{Paine2020}. This local signal is encoded mainly in a dipole. In general, the dipole has three expected contributions: a redshift independent signal from the solar systems' acceleration about the galactic centre (the secular aberration drift), a redshift\,/ distance-dependent contribution from the solar system's velocity relative to the CMB rest frame with amplitude $78\;\mu \textrm{as}\,\textrm{yr}^{-1}\,\textrm{Mpc}$ \cite{Paine2020} (the secular extragalactic parallax), and a contribution from the large-scale structure in the universe of comparable amplitude to the secular parallax \cite{Hall2019}. Of these three, the secular aberration drift is expected to be by far the largest.
\newline\indent
The advent of highly precise astrometry surveys, such as Gaia \cite{Gaia2016_Main_Paper} and Very Long Baseline Interferometry (VLBI) techniques now make it possible to detect position drift, opening a new observational window into large-scale peculiar velocities. Indeed, the largest contribution to the position drift, the secular aberration drift, has been constrained with VLBI \cite{Titov2011,Xu2012,Titov2013,Truebenbach2017} and GAIA \cite{GAIA2021_Local_Acceleration}. The most precise measurement to date is $5.05\pm 0.35\;\mu \textrm{as}\,\textrm{yr}^{-1}$, obtained with Gaia EDR3 data \cite{GAIA2021_Local_Acceleration}, and it is predicted to be constrained to better than $0.1\;\mu \textrm{as}\,\textrm{yr}^{-1}$ accuracy with future Gaia data releases. 
\newline\indent
As mentioned above, the secular aberration should be independent of redshift. Both \cite{Makarov2025} and \cite{TsigkasKouvelis2025} recently investigated the lowest-order multipoles in the position drift using quasars from Gaia DR3, finding a slight evolution of the dipole with redshift. While this evolution is not yet statistically significant, such a signal would at least naively be considered in tension with the standard cosmological model. 
\newline\newline
It is currently not clear whether or not the cosmological signal can be detected with Gaia. The author of \cite{Hall2019} predicted that the amplitude of the peculiar transverse velocities can potentially be measured with $10\sigma$ significance using the final Gaia data release. On the other hand, the authors of \cite{Duncan2024} predicted that Gaia cannot reach the required S/N for a convincing detection of this signal. However, the authors of \cite{Duncan2024} concluded that the required S/N is not far away and near-future surveys should yield convincing detection.
\newline\indent
So far, theoretical predictions for the position drift have been based on background FLRW spacetimes with linear perturbations. Considering only the dominant term in perturbation theory, the position drift is directly proportional to the transverse peculiar velocity on the sky. In this limit, position drift measurements thus allow us to constrain the parameter combination $Hf\sigma_8$ \cite{Duncan2024}, where $H$ is the expansion rate of the universe, $f$ the logarithmic derivative of the linear growth rate $\d\ln D_+/\d \ln a$, and $\sigma_8$ quantifies the amplitude of the matter power spectrum. The combination $Hf\sigma_8$ therefore probes both the large-scale expansion of the universe and the clustering and growth of matter perturbations, thereby promising insights on the nature of gravity, dark matter, and dark energy. 
\newline\indent
However, the real universe contains highly non-linear structures and in addition, observations are made on the past light cone, which is usually not taken into account in position drift studies. Neglecting this may lead to biases in theoretical expectations. For instance, it cannot be excluded that the redshift dependent signals identified in \cite{Makarov2025} and \cite{TsigkasKouvelis2025} are somehow due to observer-dependent large-scale structure effects. Given the valuable insights about our cosmos offered by the position drift and the prospect of near-future detection, it is now timely to investigate these possible biases.
\newline\indent
In order to establish the accuracy of linear perturbation theory, and to assess the effects of the non-linear structures and potential biases in our observations due to relativistic effects and large-scale structures, we here study the position drift in a fully relativistic N-body simulation run with the \texttt{gevolution} \cite{Adamek2016_Nat} code. 

This work is organised as follows. In Sect.~\ref{sect:2}, we describe our simulation setup. Sect.~\ref{sect:3} provides the necessary theoretical background on position drift, including its decomposition into spherical harmonics, the introduction of E- and B-mode power spectra, the contributions of different components of the velocity field to these spectra, and results in linear perturbation theory. In Sect.~\ref{sect:4}, we present and discuss our simulation results, comparing them with the theoretical predictions. Finally, we summarise our findings in Sect.~\ref{sect:5}.

\section{Simulations}\label{sect:2}
The simulation studied in this paper was run with the relativistic N-body code \texttt{gevolution}\footnote{\url{https://github.com/gevolution-code}}\cite{Adamek2016_Nat}. \texttt{gevolution} implements the perturbed FLRW metric in the Poisson gauge, with the line element written as
\begin{align}
	\d s^2 = a^2(\tau)\big[&-\e^{2\Psi}\,\d\tau^2-2B_i\,\d x^i\d\tau \\ \nonumber 
	&+\big(\e^{-2\Phi}\,\delta_{ij}+h_{ij}\big)\,\d x^i \d x^j \big]\;,
\end{align}
where $a$ is the background scale factor, $x^i$ are co-moving coordinates and $\tau$ is conformal time. $\Psi$ and $\Phi$ are the two scalar perturbations, known as the Bardeen potentials, while $B_i$ and $h_{ij}$ are the vector and tensor perturbations, respectively. Employing a particle-mesh scheme, the metric is evolved on a regular Cartesian grid according to a weak-field expansion of Einstein's field equations. Particle positions are updated according to the geodesic equation. For full details on the code and the expansion scheme, the reader is referred to \cite{Adamek2016}. 

\texttt{gevolution} can, while running, output particle positions and metric values on the light cone in the background spacetime for a specified observer. These outputs can then be post-processed with a ray tracer to reconstruct the exact observed light cone. For this, we use the ray tracer introduced in \cite{Adamek2019} which shoots light rays from the observer to the source in an iterative procedure. The algorithm starts by shooting a light ray towards the position of the source in the background spacetime, solving the perturbed geodesic equation along the path. Since the light ray gets deflected by structures along the way, it will initially miss the source. By adjusting the initial shooting angle and repeating the procedure, the code quickly identifies the correct null-geodesic connecting the source and observer.

To study position drift, we simulate two different light cones for a given observer: one at observation time $\tau_1$ and one at observation time $\tau_2$. The two observation times are chosen to be close to each other on cosmological time scales. The two components of the position drift are then calculated by subtracting the observed angles at the two time stamps for each source appearing on both light cones. Specifically, the position drift is calculated as 
\begin{align}
    \frac{\delta e^\theta}{\delta t} &= \frac{\theta(\tau_1)-\theta(\tau_2)}{t(\tau_1)-t(\tau_2)}\;, \label{eq:e_theta_sim} \\ 
    \frac{\delta e^\varphi}{\delta t} &= \sin(\theta(\tau_1))\,\frac{\varphi(\tau_1)-\varphi(\tau_2)}{t(\tau_1)-t(\tau_2)}\;, \label{eq:e_phi_sim}
\end{align}
where the factor $\sin(\theta)$ appears to ensure $\delta e^\varphi$ is a proper distance on the sphere. Here $\delta e^\theta,\delta e^\varphi$ refer to the $\theta,\varphi$ components of the position drift in spherical polar coordinates.

Each light cone consists of two sections: a full-sky section with radius $0.4\times \mathrm{boxsize}$ connected to a pencil beam with radius $1.05\times\mathrm{boxsize}$ and an opening half-angle of $25^\circ$. This allows maximal light-cone size without using any part of the simulation volume twice. We consider a total of ten observers with random spatial positions, each looking along one of the eight possible random box diagonals. Periodic boundary conditions are used when constructing the cones. We make sure that there is at least one proper simulation time step between $\tau_1$ and $\tau_2$. 

We use the same simulation data generated for \cite{Oestreicher2025}. The simulation contains $1024^3$ particles and has a co-moving boxsize of $1024$ Mpc/h. The metric is sampled on $1024^3$ grid cells. Since the position drift is a small effect, a low Courant factor has to be used to avoid
numerical artefacts. A Courant factor of $C=\Delta\tau/\Delta x=3$ was found to be sufficient for our purposes. For the light cones, we chose $\tau_1=\tau(z=0)$ and $\tau_2=\tau(z=0.002)$, which is the smallest possible separation with one simulation time step between the two times. 

The simulation background is a flat $\Lambda$CDM model and initial conditions are created with \texttt{CLASS}\footnote{\url{http://class-code.net}}\ \cite{CLASS_II}. The simulation used the default \texttt{gevolution} parameters: $h = H_0 / (100 \,\mathrm{km}\,\mathrm{s}^{-1}\mathrm{Mpc}^{-1}) = 0.67556,\; \Omega_{c}h^2= 0.12038,\; \Omega_{b}h^2= 0.022032,\; A_s = 2.215\times 10^{-9}$ (at $0.05\,\mathrm{Mpc}^{-1}$) and $n_s= 0.9619$. These parameters are very close to the standard Planck parameters \cite{Planck2018_Parameters}.

The simulation particles have a mass of $m \approx 1.28\times 10^{11}\,M_\odot$, making them roughly an order of magnitude lighter than the Milky Way.

\section{Theoretical Background}\label{sect:3}
In this section, we first derive expressions for the position drift in linear perturbation theory. We then show how it can be decomposed into spherical harmonics and introduce corresponding angular power spectra, showing how they relate to the three-dimensional velocity field and giving results in linear perturbation theory. We will later compare these analytical expressions to our simulation results. The analytical expression for the position drift has previously been derived in \cite{Marcori2018}. We repeat the main steps of the calculation here in order to make this paper self-contained and to introduce our notation. We consider only the dominant contributions to the position drift coming from peculiar motion and neglect contributions from metric fluctuations. This means that the wave vector is that of an FLRW spacetime and light paths and their null tangent vectors are given in accordance with an FLRW spacetime with line element 
\begin{equation}
	\d s^2 = -\d t^2 + \delta_{ij}\; a^2 \d x^i \d x^j\;,
\end{equation}
where $t$ is the cosmic time, $a$ the scale factor, $x^i$ are co-moving coordinates, and we have set $c=1$ for convenience.
\\ \\
The wave vector $k^\mu$ can be decomposed into a part perpendicular to the 4-velocity $u^\mu$ and a unit direction vector $\hat e ^\mu$ as follows
\begin{equation}
    k^\mu = \omega_\mathrm{obs}(u^\mu-\hat e^\mu)\;.
\end{equation}
Here $u^\mu = \d x^\mu /\d t$ is the peculiar velocity w.r.t. the Hubble flow, $\hat e^\mu$ fulfils $\hat e_\mu \hat e^\mu = 1$, $\hat e_\mu u^\mu = 0$, and the observed frequency is  given by
\begin{equation}
 	\omega_\mathrm{obs}= -u_\mu k^\mu\;. 
 \end{equation}
For an observer with 4-velocity $u^\mu=(1,u^i)$, the observed frequency can be written as
\begin{equation}\label{eq:w_observed}
	\omega_\mathrm{obs}=\omega (1-u_i\hat k^i)\;, 
\end{equation}
where we used that the wave vector can be written as $k^\mu = \omega (1, \hat k^i)$ in the geometric optics approximation \cite{Gravitation}. Here, $\omega$ is the angular frequency measured by a local Lorentz observer and $\hat k^i$ is a unit vector pointing in the propagation direction of the emitted light. Introducing the re-scaled wave vector $n^\mu = a/\omega\,k^\mu$, the co-moving 4-velocity $v^\mu = a u^\mu$ and $e^\mu = a \hat e^\mu$ we can write 
\begin{equation}\label{eq:n_decomposition}
	n^\mu = \frac{\omega_\mathrm{obs}}{\omega}\left(v^\mu-e^\mu\right)\;.
\end{equation}
Combining \eqref{eq:n_decomposition} and \eqref{eq:w_observed} one can show that 
\begin{equation}\label{eq:e}
	e^i = -n^i + \perp_j^i v^j\;,
\end{equation}
where
\begin{equation}
	\perp_j^i = \delta^i_j-a^{-2}n^i n_j
\end{equation}
projects orthogonally to the photon propagation direction. In words, the observed direction of a source on the sky is minus the direction in which the light was emitted, plus a correction from the observer's peculiar velocity.

%We introduce the co-moving wave vector $n^\mu = ak^\mu$ and decompose it into the co-moving 4-velocity $v^\mu = a u^\mu$ and a spatial direction 4-vector $e^\mu$ according to
%\begin{equation}\label{eq:n_decomposition}
%	n^\mu = \omega_\mathrm{obs}\left(v^\mu-e^\mu\right)\;,
%\end{equation}
%where the minus sign is introduced so that $e^\mu$ points from the observer to the source. The observed frequency is defined via
% \begin{equation}
% 	\omega_\mathrm{obs}= -u_\mu k^\mu\;.
% \end{equation}
%For an observer with 4-velocity $u^\mu=(1,u^i)$, the observed frequency can be written as
%\begin{equation}\label{eq:w_observed}
%	\omega_\mathrm{obs}=\omega (1-u_i\hat k^i)\;, 
%\end{equation}
%where $u^i=\d x^i/\d t$ is the peculiar velocity with respect to the Hubble flow and we used that the wave vector can be written as $k^\mu = \omega (1, \hat k^i)$ in the geometric optics approximation \cite{Gravitation}. Here, $\omega$ is the angular frequency measured by a local Lorentz observer and $\hat k^i$ is a unit vector pointing in the propagation direction of the emitted light. Combining \eqref{eq:n_decomposition} and \eqref{eq:w_observed} one can show that 
%\begin{equation}\label{eq:e}
%	e^i = -n^i + \perp_j^i v^j\;,
%\end{equation}
%where
%\begin{equation}
%	\perp_j^i = \delta^i_j-n^i n_j
%\end{equation}
%projects orthogonally to the photon propagation direction. In words, the observed direction of a source on the sky is minus the direction in which the light was emitted, plus a correction from the observer's peculiar velocity.

\subsection{Position Drift}
We wish to calculate the drift of a source's observed position on the sky defined via 
\begin{equation}
	\delta e^i = \frac{\d e^i}{\d t}\delta t_\o = e^i(t+\delta t)-e^i(t)\;,
\end{equation}
where $t$ is the proper time. To derive an expression for this position drift, we first note that the re-scaled wave vector $n^i$ we introduced is constant along null geodesics in an FLRW spacetime, i.e.
\begin{equation}
	n^i = \text{const.} = n_\o^i\;. 
\end{equation}
This follows from the fact that both $a^2k^i$ and $a\omega$ are constant along null geodesics in an FLRW spacetime, $n^i= a^2 k^i / (a\omega)$. 
We can then rewrite
\begin{equation}
	n^i = \frac{a}{\omega}k^i= \frac{a}{\omega}\frac{\d t}{\ d \lambda}\frac{\d\tau}{\d t}\frac{\d x^i}{\d \tau}=\frac{\d x^i}{\d \tau}
\end{equation}
and use \eqref{eq:e} to find 
\begin{equation}
	\frac{\d x^i}{\d \tau}=-e_\o^i+\perp_{j\,\o}^i v_\o^j\;. 
\end{equation}
Integrating this equation, we find
\begin{equation}
	x_\s^i-x_\o^i=(\tau_\s-\tau_\o)(\perp_{j\,\o}^i v_\o^j-e_\o^i)\;.
\end{equation}
Considering a second light ray emitted and received a short time later at $t+\delta t$ we instead have
\begin{align}
	x_\s^i+&\,v_\s^i\delta\tau_\s-x_\o^i-v_\o^i\delta\tau_o \nonumber \\ 
	=&\, (\tau_\s+\delta\tau_\s-\tau_\o-\delta\tau_\o)(\perp_{j\,\o}^i (v_\o^j+{v'}_\o^i\delta\tau_\o)-e_\o^i-\delta e_\o^i)\;, 
\end{align}
where we used the fact that 
\begin{equation}
    f(t+\delta t) = f(t)+\frac{\d f}{\d t}\delta t= f(t)+ \frac{\d f}{\d\tau}\delta\tau
\end{equation}
for any function of time $f$ and that $\perp^i_j$ is constant along null geodesics. Subtracting the two expressions from each other leaves us with 
\begin{align}\label{eq:pos_drift_derivation_step}
	v_\s^i\delta\tau_\s&-v_\o^i\delta\tau_o
	=(\tau_s-\tau_\o)(\perp_{j\,\o}^i{v'}_\o^i\delta\tau_\o-\delta e^i_\o) \nonumber \\ 
	&+(\delta\tau_\s-\delta\tau_\o)(\perp_{j\,\o}^i(v_\o^j+{v'}_\o^i\delta\tau_\o)-e_\o^i-\delta e_\o^i)\;.
\end{align}
We can now use 
\begin{align}
	\frac{\delta\tau_\o}{\delta\tau_s}=\frac{\delta t_\o}{\delta t_\s}\frac{a_\s}{a_\o}=(1+z)\,\frac{a_\s}{a_\o}=[1+(\bold v_\o-\bold v_\s)\cdot \bold n]\;
\end{align}
to find 
\begin{equation}
	\delta\tau_s-\delta\tau_\o=\delta\tau_\o(\bold v_\o-\bold v_\s)\cdot \bold n\;,
\end{equation}
where we used the expression for the perturbed redshift (see, e.g. \cite{Oestreicher2025}) and introduced the notation $\bold{a}\cdot\bold{b} = \delta_{ij}a^i b^j$ for the Euclidean scalar product. Inserting this result into \eqref{eq:pos_drift_derivation_step} we find 
\begin{align}
	v^i _\s \delta\tau_\o-v^i_\o \delta\tau_\o =&\, (\tau_s-\tau_\o)(\perp_{j\,\o}^i{v'}_\o^i\delta\tau_\o-\delta e^i_\o) \nonumber \\
	&+\, \delta\tau_\o (\bold v_\o-\bold v_\s)\cdot \bold n\, (-e^i_\o-\delta e^i_\o)\;.
\end{align}
Next, we use that $\bold v\delta e^i$ is a second-order quantity since $\delta e^i$ vanishes at background level. From this we find 
\begin{align}
	\delta\tau_\o(\delta_j^i+n_je^i_\o)(v^j_\s-v^j_\o)=(\tau_\s-\tau_\o)(\perp_{j\,\o}^i{v'}_\o^j\delta\tau_\o-\delta e^i_\o)\;.
\end{align}
Finally, we use that $\bold e = -\bold n$ at zeroth order and we are free to pick $\bold n$ at any point along the null geodesic, which implies that we can replace all $\bold n$ with $\bold e_\o$. We now also set $a_\o=1$ allowing us to replace $\perp_{j\,\o}^i=1-a_\o^{-2}n_\o^i{n_\o}_j= 1-e_\o^i {e_\o}_j$. This is possible since the affected terms always involve both $\bold n$ and $\bold v$, making any corrections arising from this replacement second order. Solving for the position drift, we find
\begin{equation}
	\delta e_\o^i = \delta t_\o \left[\frac{\perp_{j\,\o}^i(v_\s^j-v_\o^j)}{(\tau_\o-\tau_s)}+\perp_{j\,\o}^i \dot v_\o^j\right]\;,
\end{equation}
where we also replaced $\delta\tau_o = \delta t_\o$ and ${v'}_\o=\dot v_\o$, since we choose $a_\o=1$. If the observer is co-moving with the expansion of the universe, this reduces to 
\begin{equation}
	\delta e_\o^i = \delta t_\o \frac{\perp_{j\,\o}^iv_\s^j}{(\tau_\o-\tau_s)}\;.
\end{equation}
From now on, we drop the subscript and only write $\delta \bold e$ for convenience. We also replace $\tau_\o-\tau_\s=r$, where $r$ is the co-moving distance to the source, leaving us with the intuitive result
\begin{equation}\label{eq:position_drift_pertb_theory}
	\frac{\delta e^i}{\delta t_\o} = \frac{\perp_{j\,\o}^iv_\s^j}{r}\;.
\end{equation}
As $\delta \bold e$ is a vector on the sphere, it is most convenient to work in spherical polar coordinates and consider the two components $\delta e^\theta(z)$ and $\delta e^\phi(z)$.

\subsection{Harmonic Decomposition and Power Spectra}
The position drift is a vector field on the sphere with two degrees of freedom. As such, it can be decomposed into a gradient and a curl part according to
\begin{equation}\label{eq:e_decomp}
	\frac{\delta e^i}{\delta t_\o} = \nabla^i\, \Omega + \varepsilon^i_j \nabla^j\, \Psi\;.
\end{equation}
Here, $\Omega$ and $\Psi$ are scalar fields on the sphere, $\nabla$ is the covariant derivative on the sphere and $\bm\varepsilon$ is the alternating tensor. Choosing a basis $(\theta,\phi)$ one finds $\nabla^i=(\p_\theta,1/\sin\theta\;\p_\phi)$. The two potentials $\Omega$ and $\Psi$ can then be expanded into spherical harmonics as
\begin{align}\label{eq:Omega_decomp}
	\Omega (\bold e,z) &= \sum_{lm}\frac{1}{\sqrt{l(l+1)}}\,\epsilon_{lm}(z)\,Y_{lm}(\bold e)\;, \\
	\Psi (\bold e,z) &= \sum_{lm}\frac{1}{\sqrt{l(l+1)}}\,\beta_{lm}\,(z)Y_{lm}(\bold e)\;. 
    \label{eq:Psi_decomp}
\end{align}
and we can define the two angular power spectra
\begin{align}
	\delta_{ll'}\delta_{mm'}\,C_l^E (z)&= \langle \epsilon_{lm}\epsilon^*_{l'm'}\rangle\;, \\
	\delta_{ll'}\delta_{mm'}\,C_l^B (z) &= \langle \beta_{lm}\beta^*_{l'm'}\rangle\;.
\end{align}
In analogy to the CMB polarisation power spectra we refer to these as E- and B-mode spectra, respectively. 

As a vector on the sphere, the position drift itself can be decomposed either into vector spherical harmonics, as introduced in \cite{MignardKlioner2012}, or into spin weighted spherical harmonics, as introduced in \cite{Hall2019}. Both formalisms are used in the literature to study the position drift or the transverse velocity field on the sky (see, e.g. \cite{GAIA2021_Local_Acceleration, Paine2020, Makarov2025, TsigkasKouvelis2025} and \cite{Hall2019,Duncan2024} for respective examples). The two formalisms lead to the same coefficients $(\epsilon_{lm}, \beta_{lm})$ and thereby also to the same power spectra. We will mainly use the spin-weighted spherical harmonics in this publication, since these are readily available in open-source numerical implementations. In relation to comparing our results with earlier work, we will, however, refer to vector spherical harmonics. 

Below, we introduce the two decompositions after which we demonstrate how the $C_l^E, _l^B$ relate to the three-dimensional velocity field and finally give results for $C_l^E, _l^B$ in linear perturbation theory. 

\subsubsection{Spin Weighted Spherical Harmonics}\label{sect:3.2.1}
Following the formalism introduced in \cite{Hall2019} we can decompose the position drift into spin-weighted spherical harmonics. To do so, we first define the complex scalars 
\begin{equation}
    \delta e_\pm = \frac{\delta e^\theta}{\delta t_\o} \pm i\; \frac{\delta e^\phi}{\delta t_\o}\;, 
\end{equation}
arising from projecting the position drift onto the basis $\bm{\hat\theta}\pm i \bm{\hat\phi}$. The scalars $\delta e_\pm$ form a spin $\pm 1$ field and can as such be decomposed using the spin-weighted spherical harmonics. The spin $\pm 1$ spherical harmonics are given by
\begin{align}
    {}_{1}Y_{lm} &= \frac{1}{\sqrt{l(l+1)}}\eth Y_{lm}\;, \\
    {}_{-1}Y_{lm} &= -\frac{1}{\sqrt{l(l+1)}}\bar\eth Y_{lm}\;,
\end{align}
where $\eth,\bar \eth$ are the spin-raising and -lowering operator, respectively. Acting on a spin-zero quantity such as $Y_{lm}$, they are given by 
\begin{align}
    \eth Y_{lm} &= -(\p_\theta + i \frac{1}{\sin\theta}\p_\phi) Y_{lm}=-(\nabla^\theta+i\nabla^\phi)Y_{lm}\;, \\
    \bar\eth Y_{lm} &= -(\p_\theta - i\frac{1}{\sin\theta}\p_\phi) Y_{lm}=-(\nabla^\theta-i\nabla^\phi)Y_{lm}\;. 
\end{align}
Expressing $\delta e_\pm$ in terms of $\Omega$ and $\Psi$ using \eqref{eq:e_decomp}, inserting \eqref{eq:Omega_decomp},\eqref{eq:Psi_decomp} and comparing with the last two equations, one finds
\begin{equation}
	\delta e_\pm = \sum_{lm} (\mp\epsilon_{lm}+i\,\beta_{lm}){}_{\pm 1}Y_{lm}\;. 
\end{equation}
This is the decomposition we will employ when presenting our numerical results. For a detailed introduction to the spin-weighted spherical harmonics and their properties, we refer the reader to \cite{Hall2019}.

\subsubsection{Vector Spherical Harmonics}\label{sect:3.2.2}
Following \cite{MignardKlioner2012} the position drift may also be decomposed as 
\begin{equation}
    \frac{\delta \bold e}{\delta t_\o} = \sum_{lm}\left(s_{lm}\bold S_{lm} + t_{lm}\bold T_{lm}\right)\;. 
\end{equation}
where $\bold S_{lm}$ and $\bold T_{lm}$ are the vector spherical harmonics defined as 
\begin{align}
    \bold S_{lm} &= \frac{1}{\sqrt{l(l+1)}}\,r\,\nabla_{\bold r}Y_{lm}\;, \\
    \bold T_{lm} &= -\bold{\hat r} \times \bold S_{lm}\;.
\end{align}
where $\nabla_{\bold r}$ is now the 3-dimensional gradient operator. Here $\bold r$ is the radial vector in spherical polar coordinates pointing from the observer to the source, $r$ is its absolute value, the co-moving distance introduced above, and $\hat{\bold r}=\bold r /r$ is the corresponding unit vector. It is straightforward to show that this is equivalent to 
\begin{align}
    S_{lm}^i &= \frac{1}{\sqrt{l(l+1)}}\,\nabla^iY_{lm}\;, \\
    T_{lm}^i &= \frac{1}{\sqrt{l(l+1)}}\varepsilon^i_j \nabla^i Y_{lm}\;. 
\end{align}
Inserting \eqref{eq:Omega_decomp},\eqref{eq:Psi_decomp} into \eqref{eq:e_decomp} and identifying $S^i_{lm},T^i_{lm}$, one immediately sees that $(s_{lm},t_{lm})=(\epsilon_{lm},\beta_{lm})$.

Naturally, the exact coefficients and vector spherical harmonics depend on the chosen coordinate system. Here, we work with the usual spherical polar coordinates with angles $(\theta,\phi)$. However, in the astronomy literature, it is very common to work with equatorial coordinates with angles $(\delta, \alpha)$. Since the two coordinate systems have the same reference axis, the $Y_{lm}$ remain unchanged when moving between these two coordinate systems. They therefore lead to the exact same coefficients $(s_{lm},t_{lm})$. We are thus free to work in either coordinate system and can still directly compare our obtained multipole coefficients obtained using $(\theta,\phi)$ with results obtained using $(\delta, \alpha)$. 
\\\\
As detailed in Sect.~4 of \cite{MignardKlioner2012} a vector field on the sphere has two global features: a rotation described by a vector $\bold R$, and a glide described by a vector $\bold G$. The rotation generates the vector field 
\begin{equation}
    \bold v^R = \bold R \times \bold{\hat r}  
\end{equation}
while the glide generates the vector field
\begin{equation}
    \bold v^G = \bold{\hat r}  \times (\bold G \times \bold{\hat r} )=\bold G-(\bold G\cdot \bold{\hat r} )\bold{\hat r} \;.
\end{equation}
In terms of the vector spherical harmonics one simply has 
\begin{align}
    \bold v^R = \sum_m t_{1m}\bold T_{1m}\;, \qquad \bold v^G = \sum_m s_{1m}\bold S_{1m}\;,
\end{align}
i.e. a global rotation contributes only to the B-mode and a global glide only to the E-mode. For the individual components, one explicitly has 
\begin{align}
    t_{10}&=\sqrt{\frac{8\pi}{3}}R_3\;, \quad t_{11} = -t_{1,-1}^* = \sqrt{\frac{4\pi}{3}}(-R_1+i R_2)\;, \nonumber \\ 
    s_{10}&=\sqrt{\frac{8\pi}{3}}G_3\;, \quad s_{11} = -s_{1,-1}^* = \sqrt{\frac{4\pi}{3}}(-G_1+i G_2)\;.
\end{align}
Using that the $C_l$ are defined as
\begin{equation}
    C_l^B = \frac{1}{2l+1}\sum_{m=-l}^l \beta_{lm}\beta_{lm}^*\;, \qquad C_l^E = \frac{1}{2l+1}\sum_{m=-l}^l \epsilon_{lm}\epsilon_{lm}^*
\end{equation}
and remembering that $(s_{lm},t_{lm})=(\epsilon_{lm},\beta_{lm})$ we can find the amplitude of the rotation and glide simply from 
\begin{align} 
    R = \sqrt{\frac{9}{8\pi}C_1^B}\;, \qquad G = \sqrt{\frac{9}{8\pi}C_1^E}\;.
    \label{eq:rot_glide_amp}
\end{align}
The amplitude of the glide is equivalent to the amplitude of the dipole in the position drift signal. We later constrain the contribution from the large-scale structure to this dipole. The rotation is only non-zero in spacetimes with intrinsic vorticity and has no significant contribution from large scale structure.

\subsubsection{Theoretical Power Spectra}
We now wish to understand how the two angular power spectra $C_l^E, C_l^B$ for the position drift are related to the cosmological velocity field. The velocity field can be decomposed into a gradient and rotational part according to
\begin{equation}
	\bold v  = -\nabla_{\bold r}\, v + \bold v_R\;,
\end{equation}
where $\nabla_{\bold r} \cdot \bold v_R = 0$. Switching to spherical polar coordinates, the rotational part, $\bold v_R$, can be further decomposed into a toroidal and poloidal part 
\begin{equation}\label{eq:v_decomp_r}
	\bold v = - \nabla_{\bold r}\, v + \nabla_{\bold r} \times (v_T \bold{\hat r}) + \nabla_{\bold r}\times(\nabla_{\bold r} \times (v_P \bold{\hat r}))\;,
\end{equation}
where $v_T$ and $v_P$ are scalar potentials (see, e.g. Appendix III in \cite{Chandrasekhar1961}). Projecting the velocity onto the sphere one then finds 
\begin{equation}
	v_\perp^i =  \frac{1}{r}\left(-\nabla^i\, v+\varepsilon^i_j\nabla^j\, v_T + \nabla^i\, \p _r v_P\right)\;.
\end{equation}
This can easily be verified by inserting the usual expression for $\nabla_{\bold r}$ in spherical polar coordinates into \eqref{eq:v_decomp_r} and evaluating the expression. Comparing to \eqref{eq:position_drift_pertb_theory} and \eqref{eq:e_decomp} we can now read off 
\begin{align}
	\Omega (\bold e, z)&= \frac{1}{r^2}(-v+\p_r v_P)\;, \label{eq:Omega} \\
	\Psi (\bold e, z)&=\frac{1}{r^2}\,v_T\;. \label{eq:Psi}
\end{align}
We now focus on the E-mode power spectrum. Expanding the contributions to $\Omega$ into Fourier modes, we have
\begin{align}
	\Omega (\bold e, z)= \frac{1}{r^2}\Bigg(&-\int\frac{\d^3 k }{(2\pi)^3}v(\bold k ) \e^{-ir\bold k \cdot \bold e}\nonumber \\
    &+\p_r\int\frac{\d^3 k }{(2\pi)^3}v_P(\bold k ) \e^{-ir\bold k \cdot \bold e}\Bigg)\;.
\end{align}
Using the identity 
\begin{equation}
	\e^{-i r \bold k\cdot\bold e} = 4\pi \sum_{l=0}^\infty\sum_{m=-l}^l (-i)^l j_l(kr) Y_{lm}(\bold e)Y_{lm}^*(\bold{\hat k})\;
\end{equation}
to expand the exponential into spherical harmonics and comparing with \eqref{eq:Omega_decomp}, we can read off that 
\begin{align}
	\epsilon_{lm}(z) =(-i)^l\frac{\sqrt{l(l+1)}}{r^2}\int\frac{\d^3 k }{2\pi^2}\Big[&-v(\bold k)j_l(kr)\nonumber\\
    &+v_P(\bold k)kj'_l(kr)\Big]Y_{lm}^*(\bold{\hat k})\;.
\end{align}
Here, $j_l$ are the spherical Bessel functions and $j_l'$ denotes the first derivative with respect to the argument.
We can now calculate the power spectrum 
\begin{align}
	\delta_{ll'}\delta_{mm'}\,C_l^E &= \langle \epsilon_{lm}\epsilon^*_{l'm'}\rangle\;.
\end{align}
Introducing power spectra $P_{xy}(k)$ defined via
\begin{equation}
	\langle x(\bold k)y^*(\bold k')\rangle = (2\pi)^3\delta_D(\bold k -\bold k')P_{xy}(k)\;, 
\end{equation}
integrating over delta functions, using the orthonormality of the spherical harmonics to solve a remaining angular integral, and using $P_{vv_P}=P_{v_P v}$ it follows that
\begin{align}
	C_l^E(z)=\frac{2}{\pi}\frac{l(l+1)}{r^4}\int\d k\, k^2 \Big[&j^2_l(kr)P_{vv}(k)+k^2{j'_l}^2(kr)P_{v_P v_P}(k)\nonumber\\ 
    &-2kj_l(kr)j'_{l}(kr)P_{v v_P}(k)\Big]\;.
\end{align}
For the B-mode the calculation follows the same steps, but comparing \eqref{eq:Omega} and \eqref{eq:Psi} we see that the result for the B-mode directly follows from the E-mode, yielding
\begin{align}
	C_l^B(z)=\frac{2}{\pi}\frac{l(l+1)}{r^4}\int\d k k^2 \Big[j^2_l(kr)P_{v_Tv_T}(k)\Big]\;.
\end{align}
The B-mode power spectrum is therefore entirely sourced by the toroidal part of the velocity field, while the E-mode power spectrum is sourced both by the usual gradient potential $v$ and the poloidal part of the velocity field. 

Calculating fully non-linear power spectra in cosmology is still an open problem due to the collisionless nature of dark matter \cite{Bernardeau2002}. In the next subsection, we therefore limit ourselves to large scales and small perturbations in the density and velocity field and accordingly give results for the two power spectra within linear perturbation theory. 

\subsubsection{Linear Perturbation Theory}
To lowest order in perturbation theory the cosmic velocity field is rotation free, meaning $\bold v_R=0$ and thereby $v_T,v_P=0$. We therefore expect 
\begin{align}
	C_l^E(z)&=\frac{2}{\pi}\frac{l(l+1)}{r^4}\int\d k\, k^2 j^2_l(kr)P_{vv}(k)\;, \\
    C_l^B(z)&= 0\;.
\end{align}
Using the linear continuity equation $a\dot\delta = -k^2v$ we can make the replacement 
\begin{equation}
	P_{vv}(k) = \left(\frac{1}{k^2}a f H D_+\right)^2 P_\delta^0(k)\;, 
\end{equation}
where $D_+$ is the linear growth factor and we introduced the growth rate $f=\d\ln D_+/\d\ln a$ and the density fluctuation power spectrum at present time $P_\delta^0$. This leaves us with 
\begin{equation}
	C_l^E(z) = \frac{2}{\pi}\frac{l(l+1)}{r^4}(afHD_+)^2\int\,\frac{\d k}{k^2}j^2_l(kr)P^0_\delta(k)\:.
\end{equation}
For ease of notation, we introduce 
\begin{equation}
	T(z)=\frac{afHD_+}{r^2}\;.
\end{equation}
When evaluating observational data, we usually do not consider data at one specific redshift, but rather all points within a small redshift window $z\pm\d z$. We account for this by adding a window function, $W(z)$, (see, e.g., \cite{CLASSgal})
\begin{equation}
	C_l^{E}(z) = \frac{2}{\pi}l(l+1)\int \frac{\d k}{k^2} P_\delta^0(k) \left(\int \d z\; W(z) j_l(kr(z))T(z)\right)^2\;.
	\label{eq:psd_Cl}
\end{equation}
We use a normalised tophat window function
\begin{equation}
	W(z) = 
	\begin{cases} 
		1/(2\d z) & \text{if} \quad z-\d z \leq z \leq z+\d z\;, \\
		0         & \text{otherwise}\;.
	\end{cases}
\end{equation}
Here $\d z$ refers to the half-width of the tophat. To solve the highly oscillatory integrals in \eqref{eq:psd_Cl}, we use the library \texttt{FFTlog-and-beyond}\footnote{\url{https://github.com/xfangcosmo/FFTLog-and-beyond}}\cite{Fang_2020_FFTlog} which was specifically designed to handle integrals over spherical Bessel functions and their derivatives. Using this library, we solve the inner integrals over $z$, writing them as functions of $r$ to do so. We then use the composite Simpson's rule to solve the remaining $k$-integral. To evaluate the linear matter power spectrum at present time, $P_\delta^0(k)$, as well as the cosmological functions, we use CLASS with the same cosmological parameters listed in Sect.~\ref{sect:2}. Our implementation can be found here \href{https://github.com/oestreichera/positiondrift_cl}{\faicon{github}}.

\begin{figure*}
	\centering
	\includegraphics[width=0.4\textwidth]{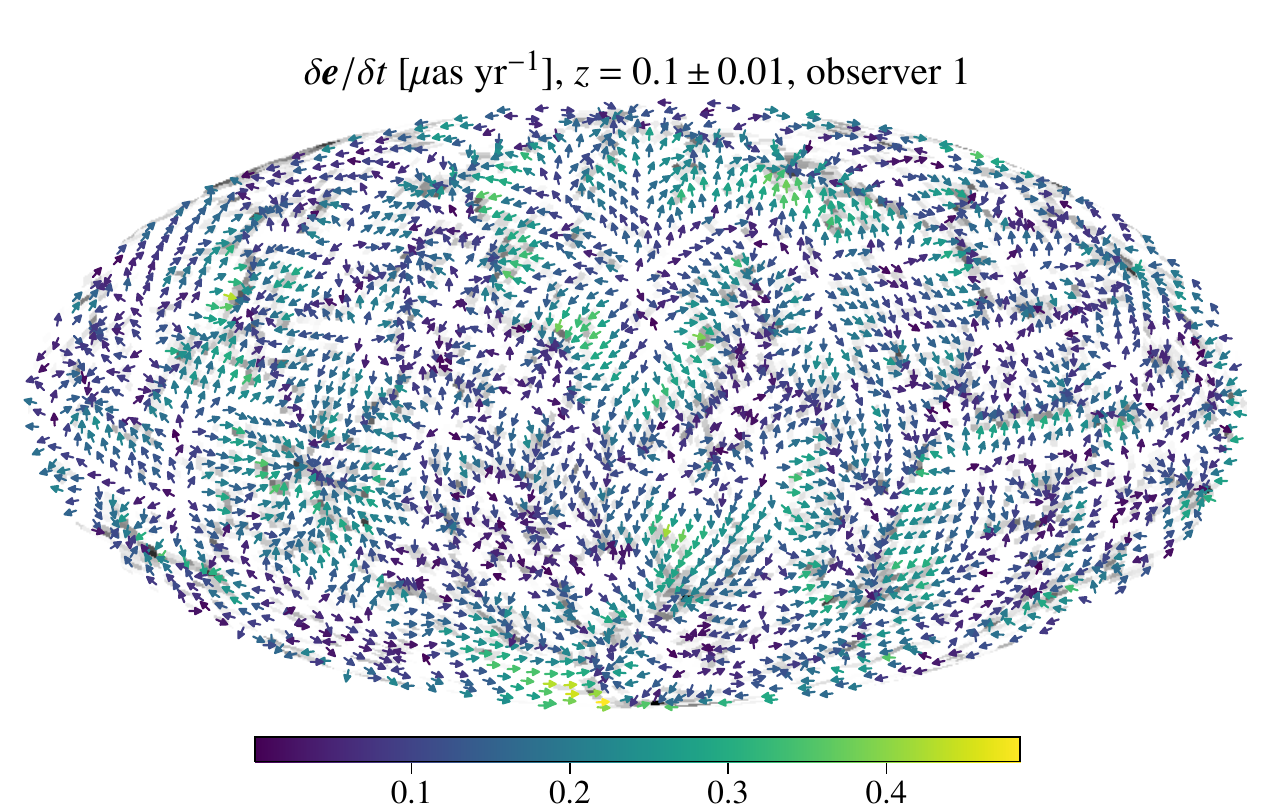}%
	\includegraphics[width=0.24\textwidth]{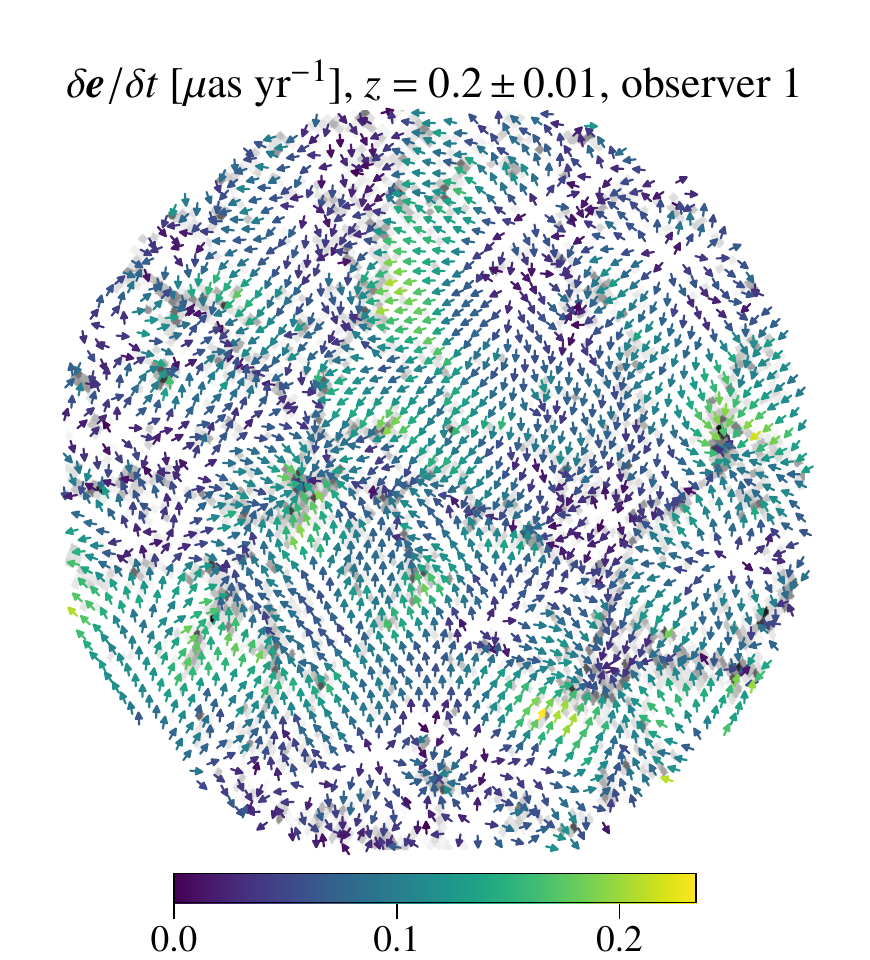}%
	\includegraphics[width=0.24\textwidth]{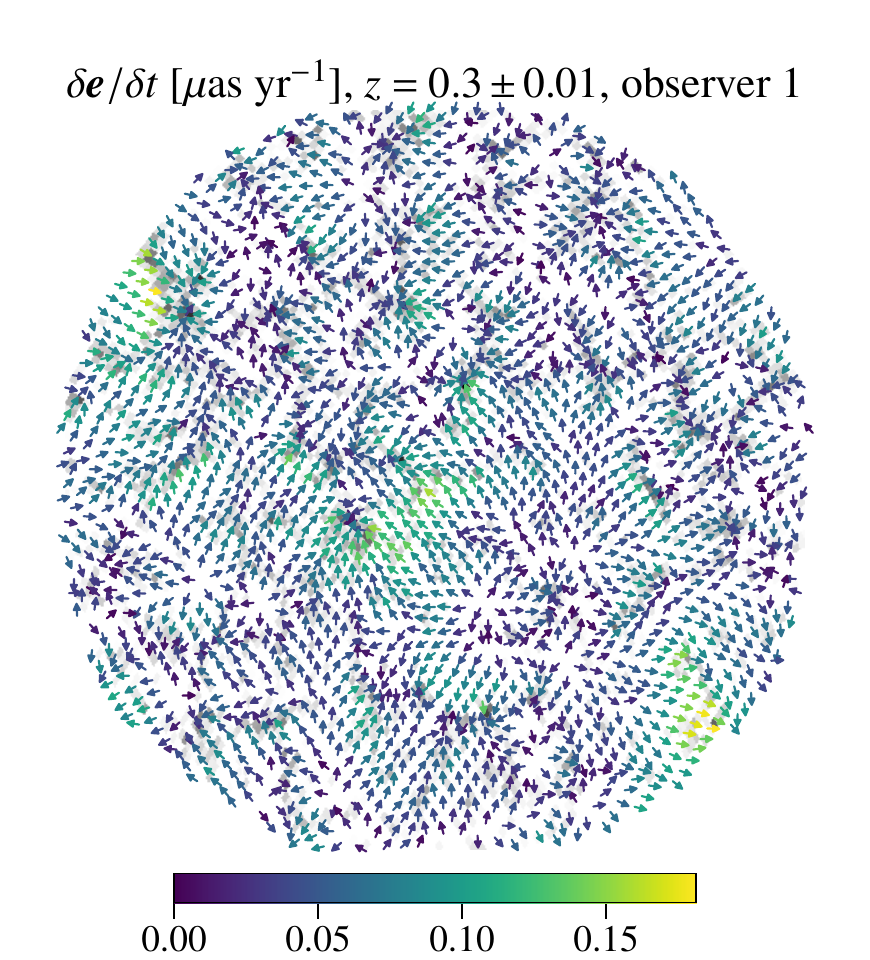}
	\includegraphics[width=0.4\textwidth]{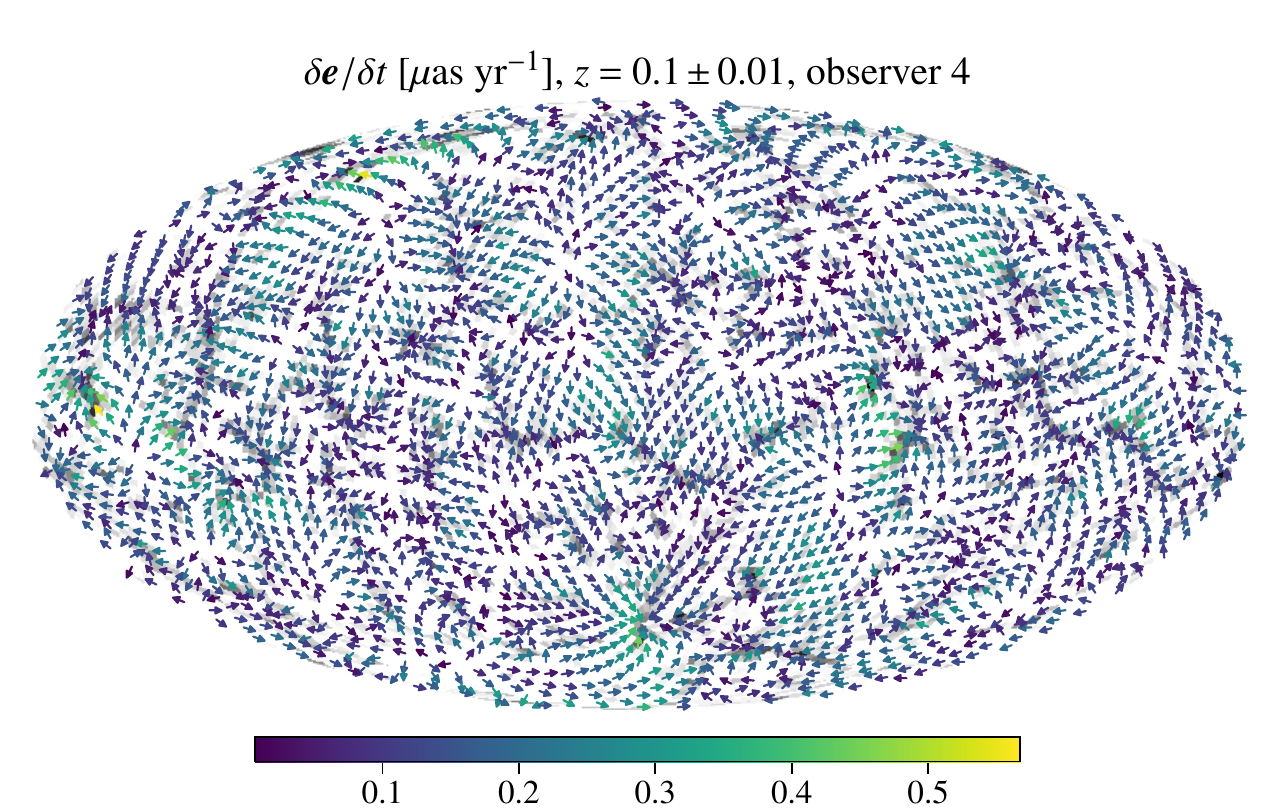}%
	\includegraphics[width=0.24\textwidth]{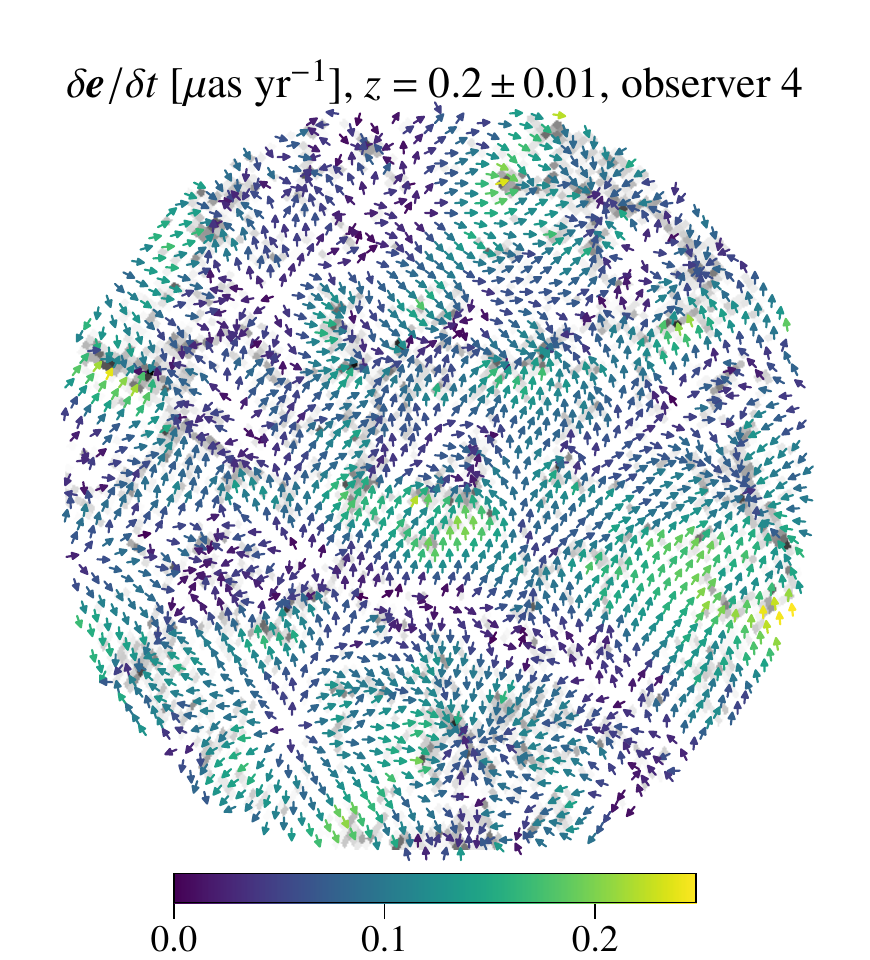}%
	\includegraphics[width=0.24\textwidth]{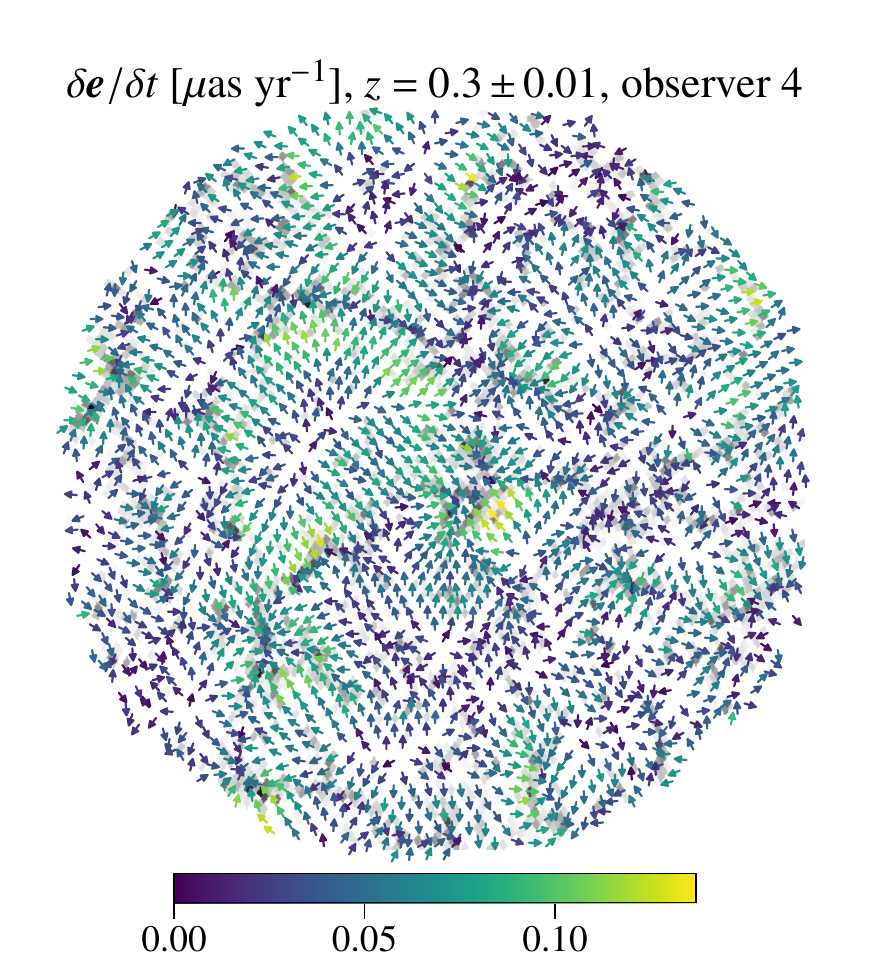}
	\includegraphics[width=0.4\textwidth]{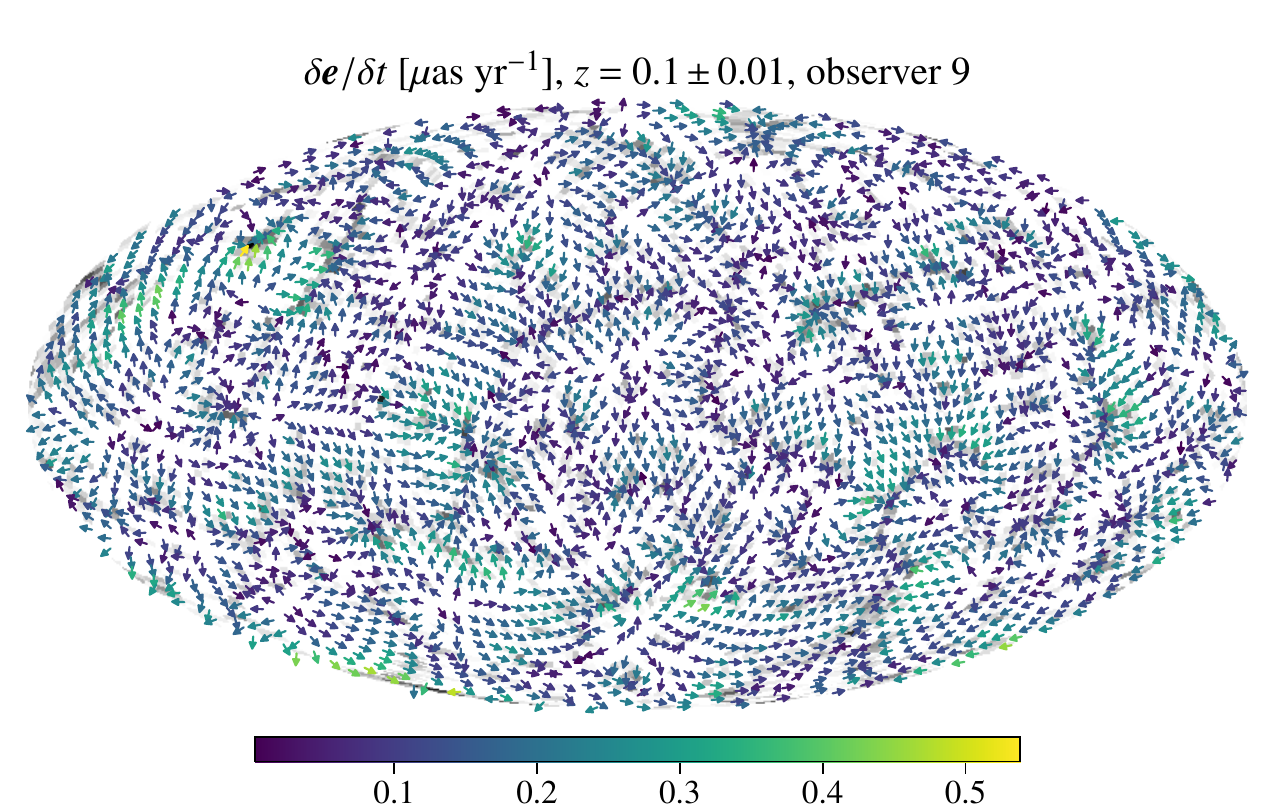}%
	\includegraphics[width=0.24\textwidth]{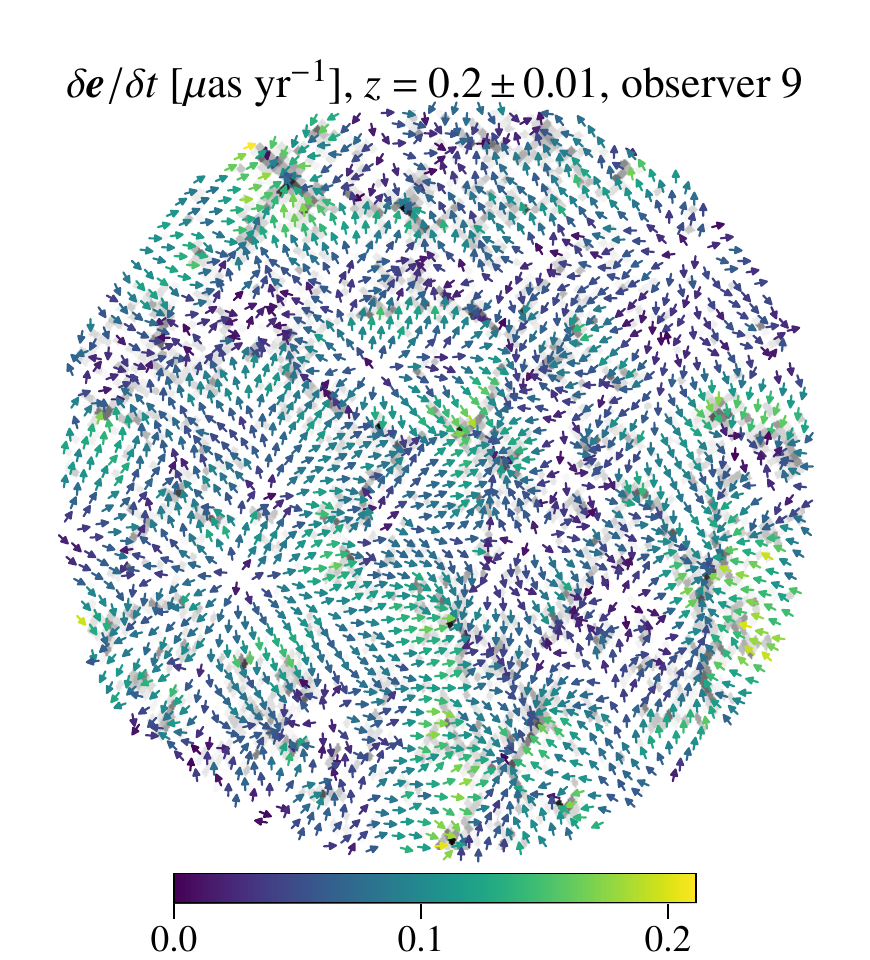}%
	\includegraphics[width=0.24\textwidth]{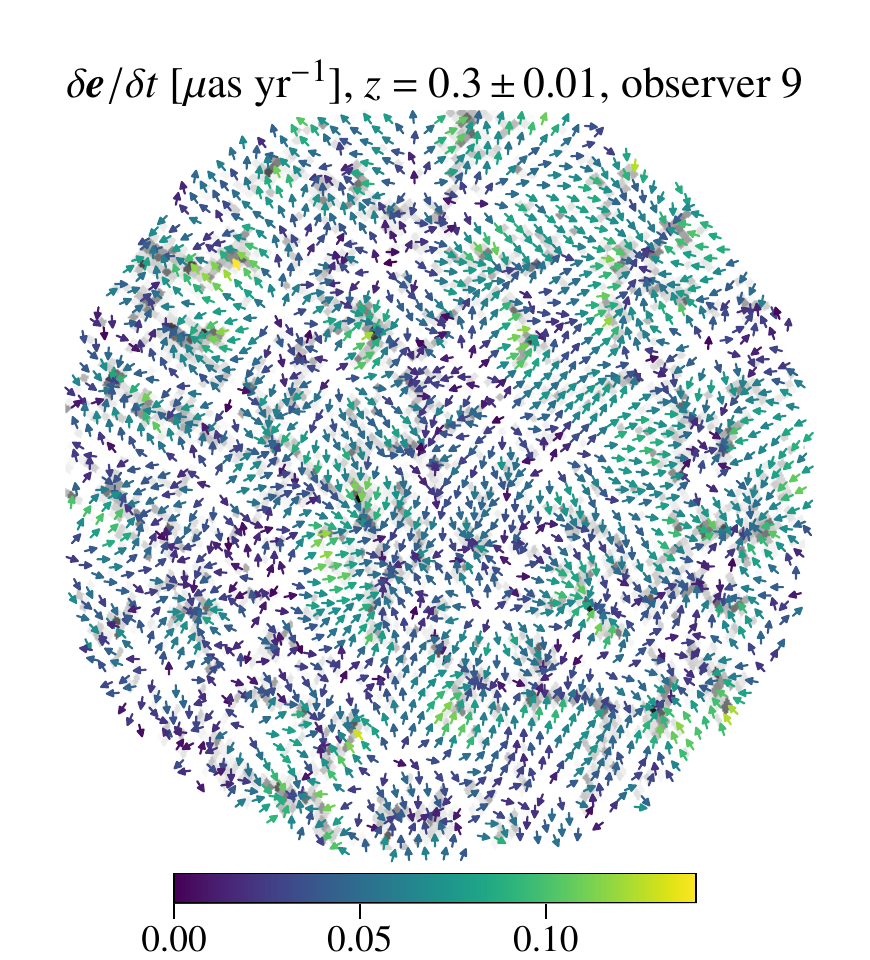}
	\caption{HEALPix skymaps of the position drift, for three different redshift bins $z=0.1,0.2,0.3\pm0.01$ and three different observers. The map for $z\sim 0.1$ shows the full sky with resolution $N_\mathrm{side}=16$, while the maps for $z\sim 0.2,0.3$ show partial circular sections of the sky from a pencil beam light cone with half-opening angle $25^\circ$ and have a resolution $N_\mathrm{side}=64$. The colour of the vectors indicates the absolute value of the position drift. The highest density regions are indicated in grey in the background, visible upon zooming in.}
    \label{fig:skymaps}
\end{figure*}

\section{Results}\label{sect:4}
In this section, we show the results of our simulation and compare them with the theoretical predictions derived in Sect.~\ref{sect:3}. To keep the data amount manageable, we show results using every one-hundredth particle in the light cone unless otherwise specified. 
\\ \\
We start by showing HEALPIX sky maps of the position drift in Fig.~\ref{fig:skymaps} for three of our observers. For the full-sky map at $z\sim 0.1$ we choose a resolution of $N_\textrm{side}=16$ and for the partial maps at $z\sim 0.2,0.3$ we choose $N_\textrm{side}=64$, averaging the position drift in cells with multiple particles. We indicate the direction of the drift with small arrows and its absolute value with the colour of the arrows. In the background, we indicate the most overdense structures in grey. As expected, the position drift points towards overdensities and away from voids. The maps show that the signal is dominated by quite large bulk flows. There are no significant differences between the different observers.

\begin{figure*}
    \centering
	\includegraphics[width=0.33\linewidth]{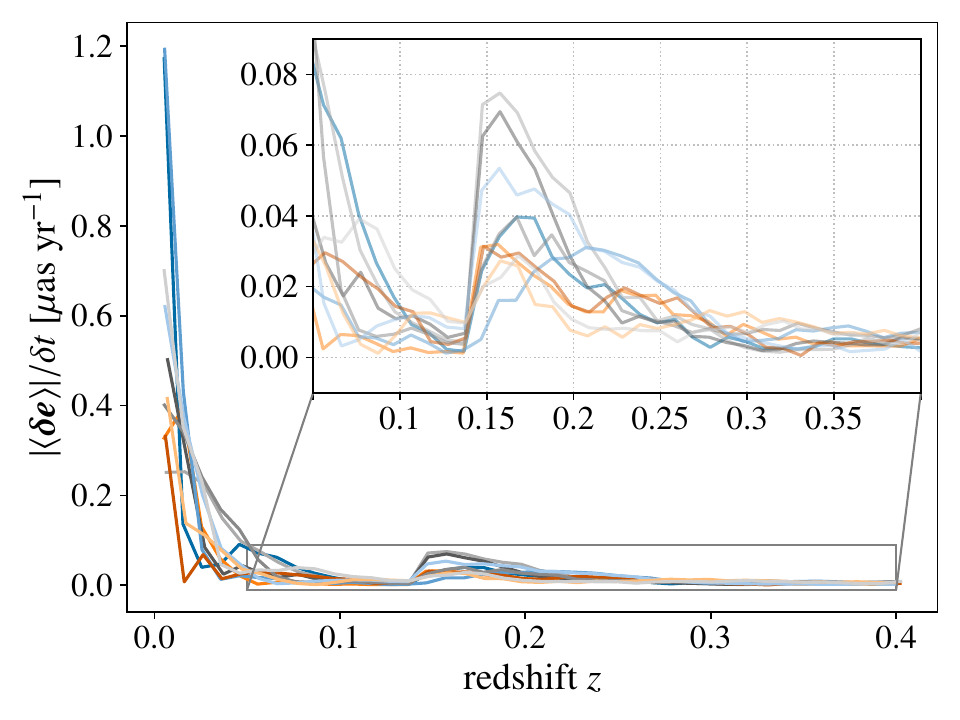}%
	\includegraphics[width=0.33\linewidth]{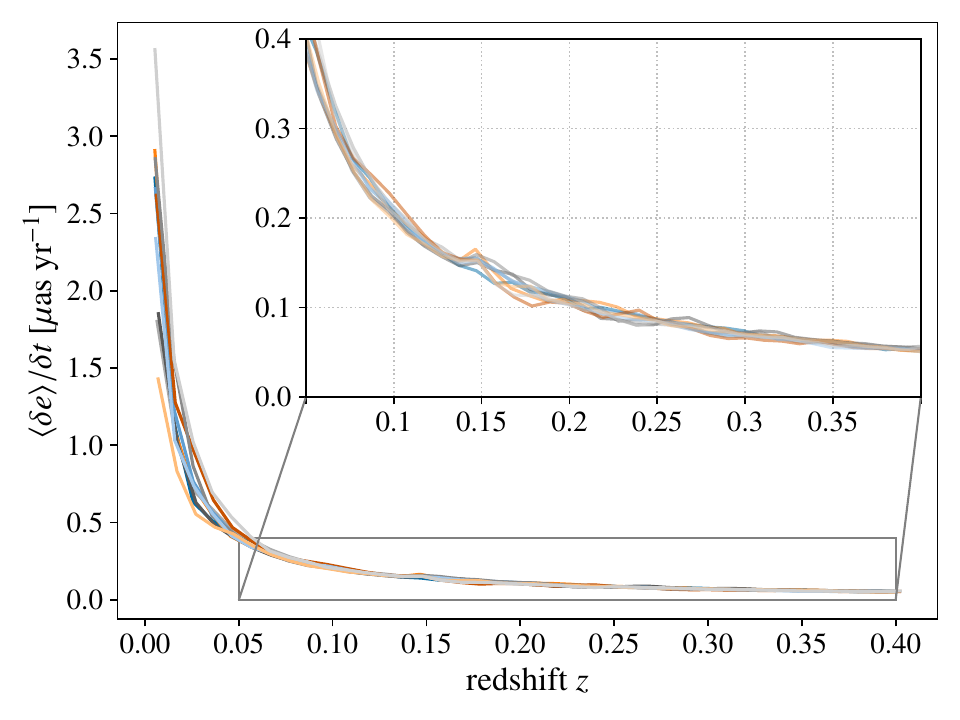}%
	\includegraphics[width=0.33\linewidth]{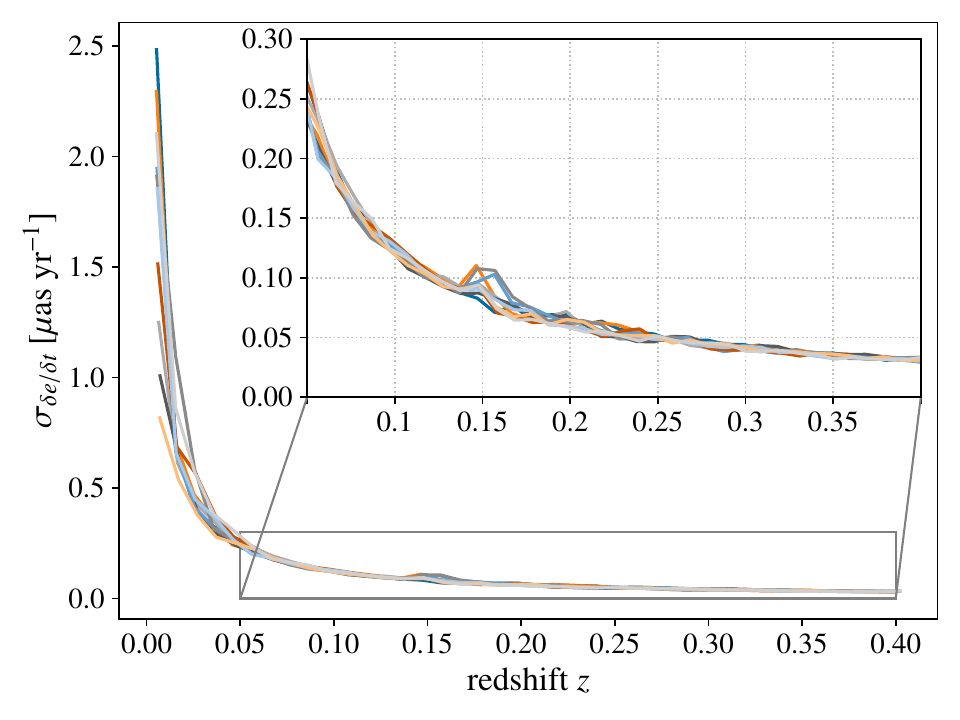}
	\caption{Absolute value of the mean (left), mean of the absolute value (centre), and standard deviation of the absolute value (right) of the position drift averaged in 40 redshift bins over the entire simulated light cone for 10 different observers. Full sky coverage up to redshift $z\sim0.15$ and partial sky coverage, i.e. a pencil beam with half opening angle $25^\circ$, for larger redshifts.}
    \label{fig:mean_std}
\end{figure*}

In Fig.~\ref{fig:mean_std}, we show the absolute value of the mean (left panel), the mean of the absolute value (centre), and the standard deviation of the absolute value of the position drift in 40 bins for the ten observers. In a perfect FLRW universe, the absolute value of the mean vanishes, but as the figure reveals, in a universe with structures, there is a local observer-dependent nonvanishing mean. The mean quickly decreases as a function of the redshift, and already at $z\approx 0.1$, the fluctuations of the mean are below $0.05\, \mu \textrm{as}\,\textrm{yr}^{-1}$ for all ten observers. One can see a small increase in the mean at redshift $z\sim0.15$. This is an artefact arising from the transition from the full-sky light cone to the partial light cone. Specifically, with only partial sky coverage, the mean converges to zero much more slowly than when considering the particles on the entire sky. We have confirmed that the transition is smooth if a pencil beam is considered for the entire redshift range. 

\begin{figure*}
	\centering
	\includegraphics[width=0.33\linewidth]{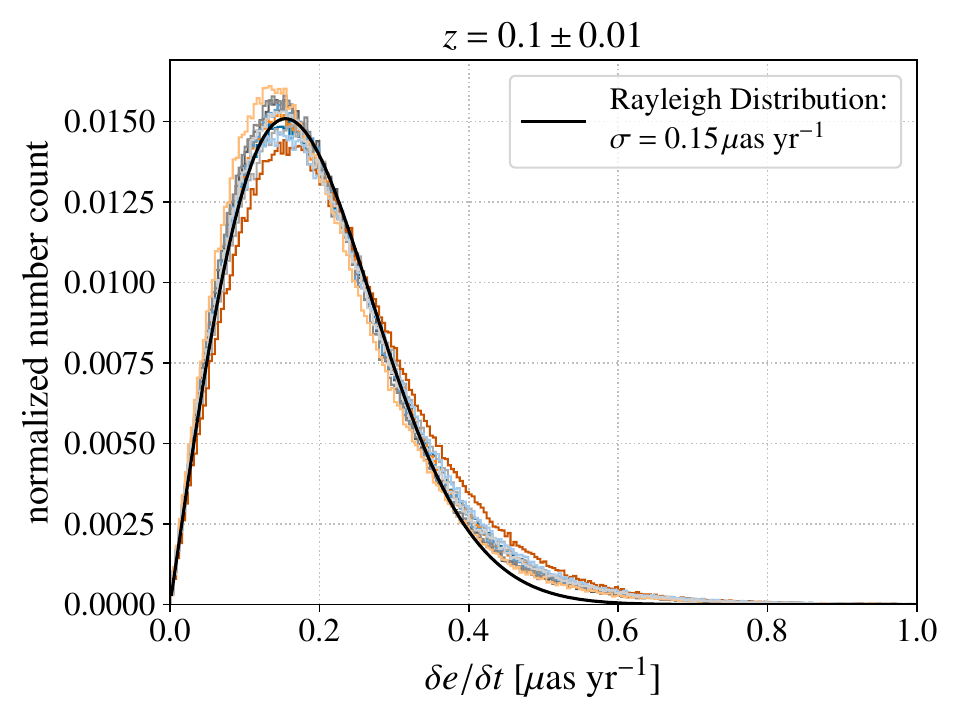}%
	\includegraphics[width=0.33\linewidth]{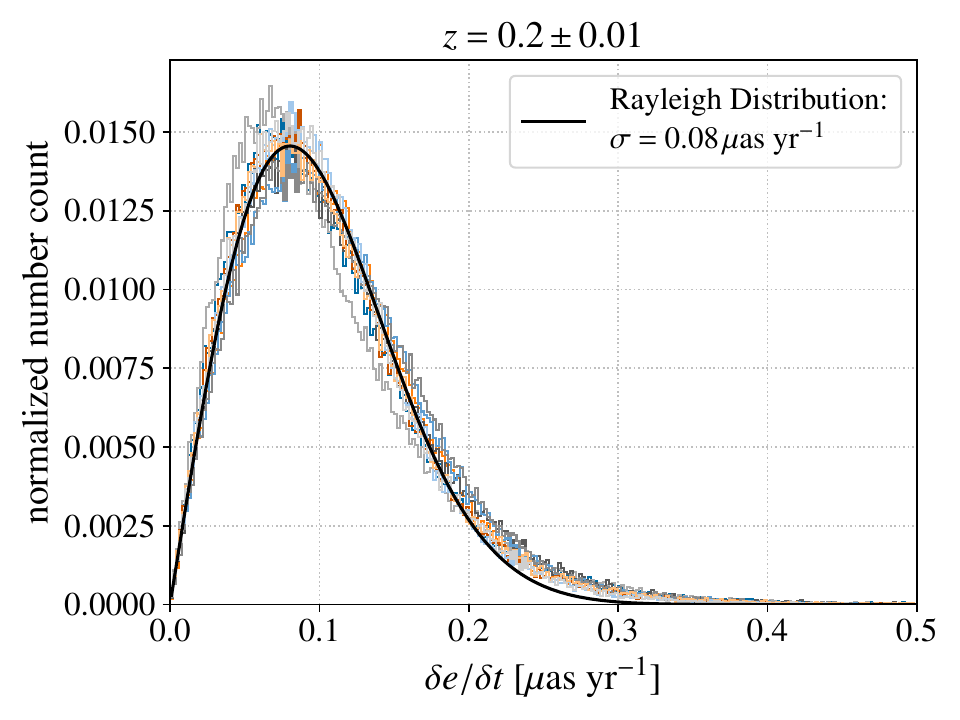}%
	\includegraphics[width=0.33\linewidth]{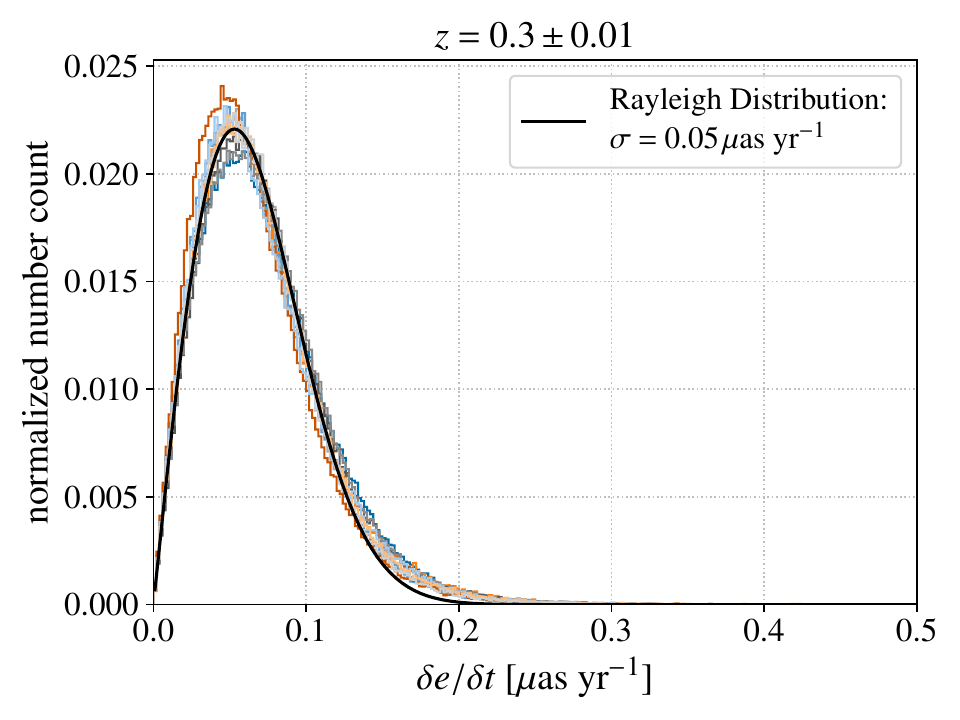}
	\caption{Normalized histograms of the absolute value of the position drift in three different redshift bins $z= 0.1,0.2,0.3\pm0.01$ for ten different observers and best-fit Rayleigh distribution \eqref{eq:rayleigh_dist} with scale parameter $\sigma$.}
    \label{fig:histograms}
\end{figure*}

Individual particles on the other hand clearly have a non-zero position drift as the centre panel demonstrates. The mean absolute value of the position drift is largest for smallest redshifts and decreases for higher redshifts, reaching $0.1\, \mu \textrm{as}\,\textrm{yr}^{-1}$ around $z\approx 0.2$ for all ten observers. The standard deviation shows similar behaviour, reaching $0.1\, \mu \textrm{as}\,\textrm{yr}^{-1}$ around $z\approx 0.15$.
\\ \\ 
In Fig.~\ref{fig:histograms} we show normalised histograms of the position drift in three different redshift bins with centres at $z= 0.1,0.2,0.3$ and widths given by $\pm0.01$. The histograms are shown for all ten observers individually. The distribution is well described by a Rayleigh distribution with scale parameter $\sigma$
\begin{equation}
    f(x) \propto \frac{x}{\sigma^2}\e^{-x^2/(2\sigma^2)}\;,
    \label{eq:rayleigh_dist}
\end{equation}
as we would expect if the individual components of the transverse velocity follow a Gaussian distribution with standard deviation $\sigma$. In Fig.~\ref{fig:histograms}, we also show the best-fit Rayleigh distribution to the mean of all observers. There is a small difference between the best fit and the data in the tail of the distribution. This could either result from the fact that the data are taken in small redshift bins rather than at a single exact redshift or from non-linearities on small scales, where velocities become larger and non-Gaussian. There are only small deviations between the individual observers. 

\subsection{Comparison to Theoretical Predictions}
\begin{figure*}
	\centering
	\includegraphics[width=0.33\linewidth]{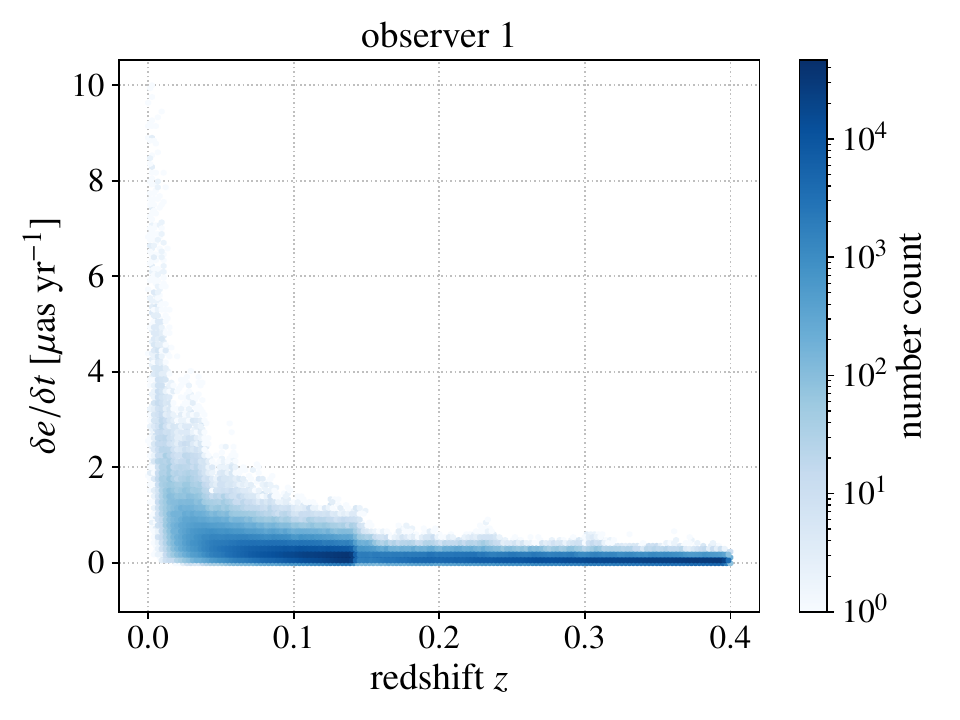}%
	\includegraphics[width=0.33\linewidth]{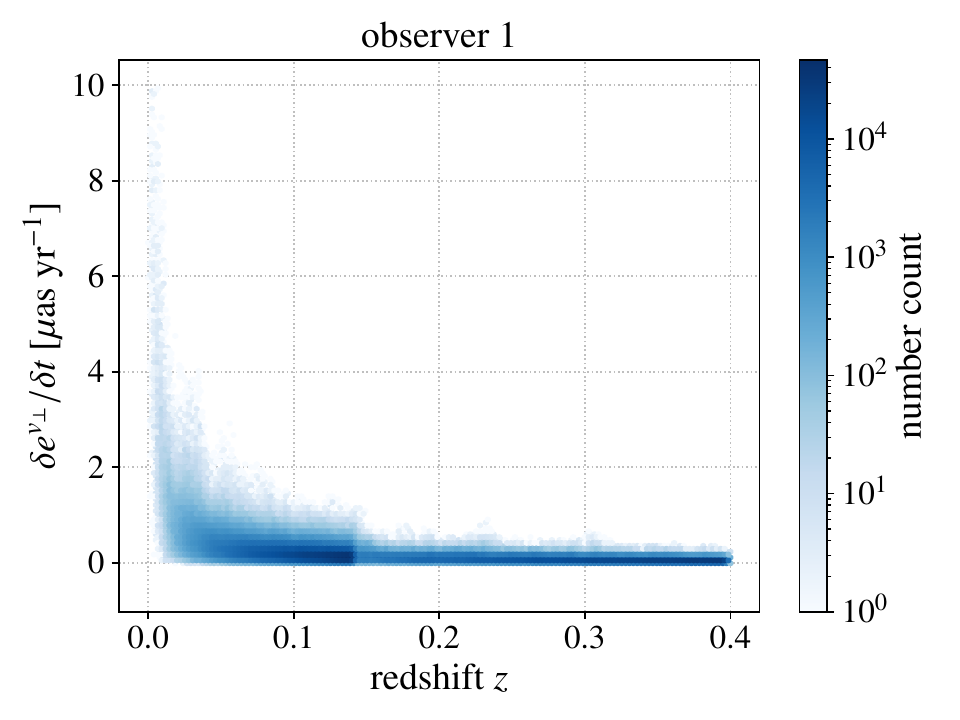}%
	\includegraphics[width=0.33\linewidth]{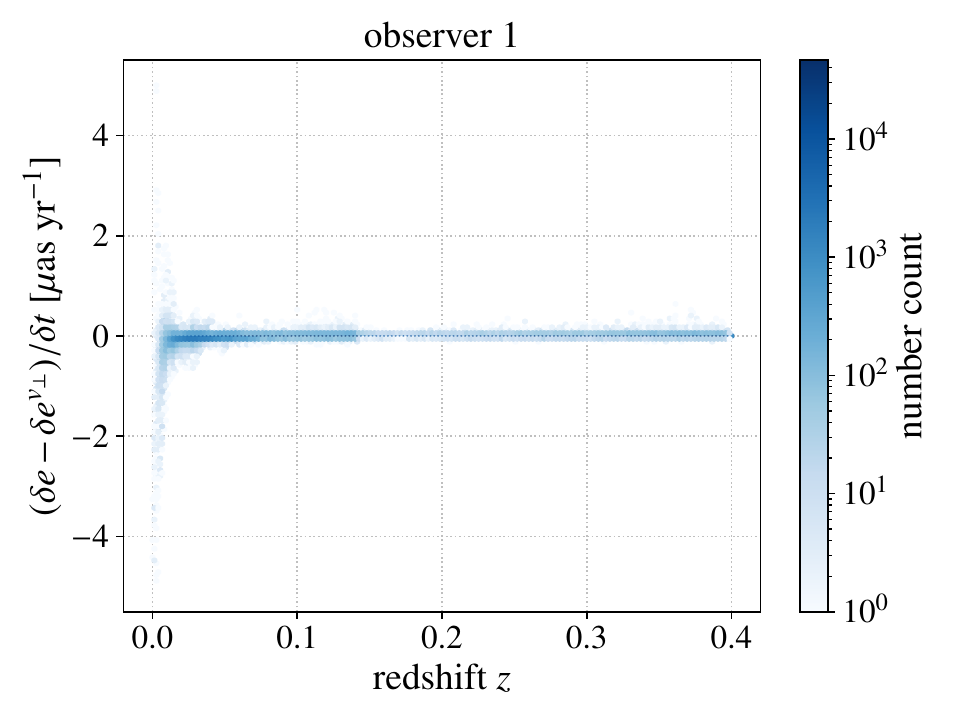}
	\caption{Hexbin density map of the absolute value of the position drift as a function of the redshift. Simulation result (left), prediction according to perturbation theory as given by \eqref{eq:position_drift_pertb_theory} (centre) and difference between the two (right).}
    \label{fig:hexbin}
\end{figure*}

In Fig.~\ref{fig:hexbin} we show hexbin density maps of the absolute value of the position drift for one of our observers. The left panel shows the result obtained from the simulation. As seen, at low redshift, the position drift can be as large as $10\,\mu\textrm{as}\,\textrm{yr}^{-1}$ for individual sources.

\begin{figure*}
	\centering
	\includegraphics[width=0.33\linewidth]{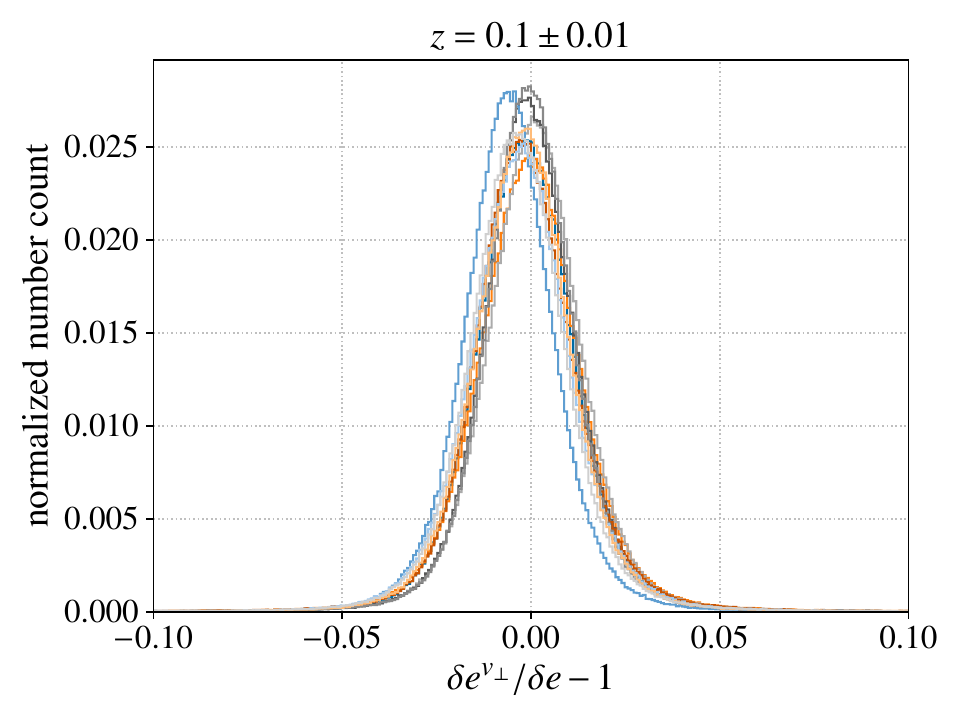}%
	\includegraphics[width=0.33\linewidth]{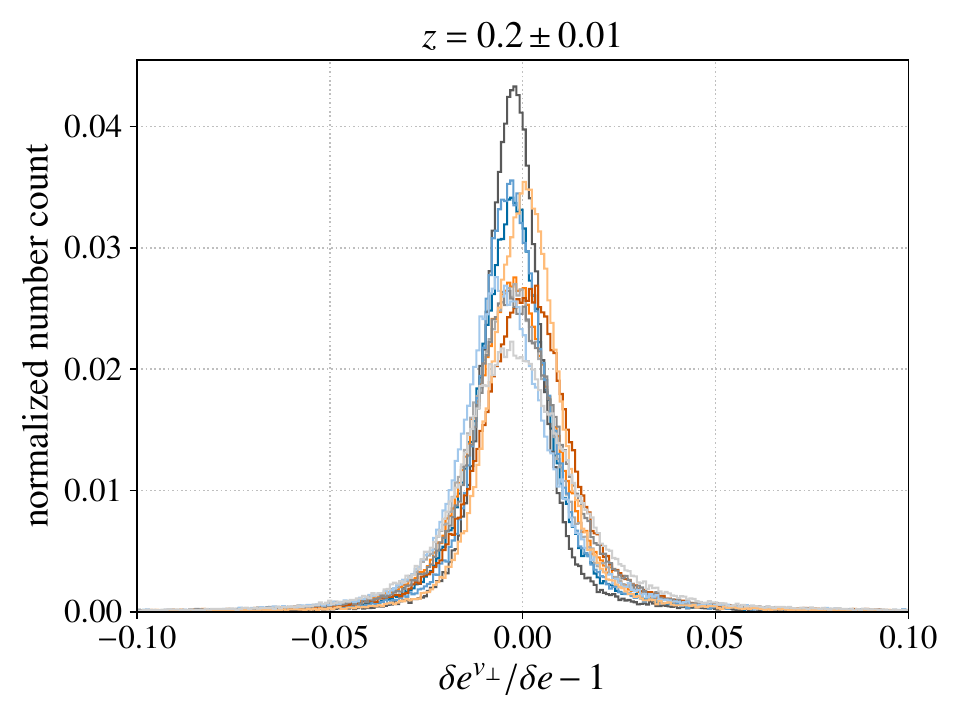}%
	\includegraphics[width=0.33\linewidth]{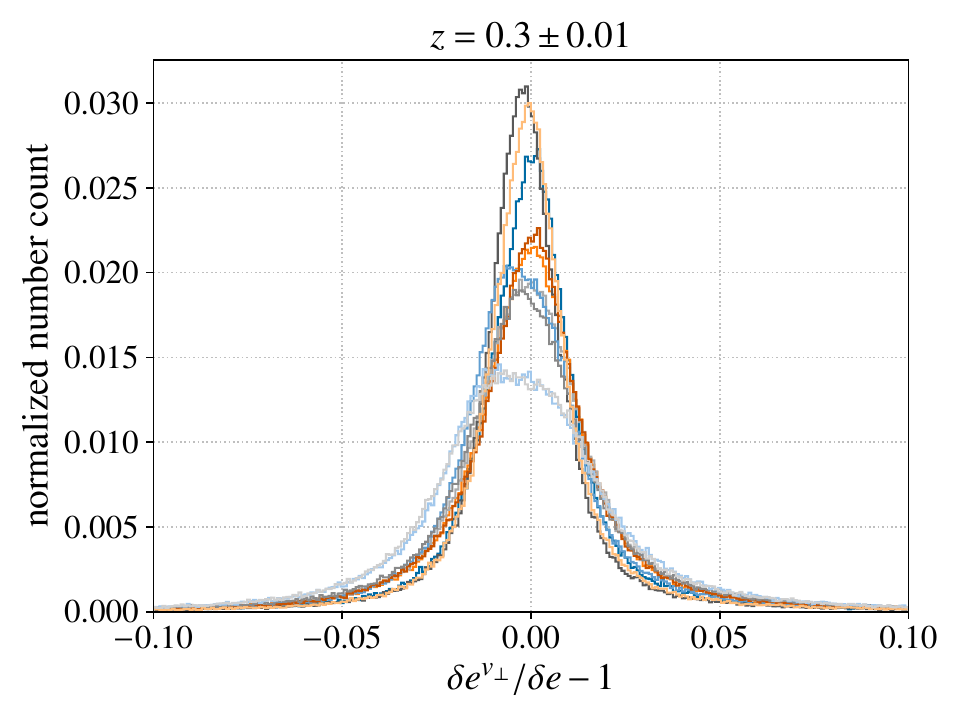}
	\includegraphics[width=0.33\linewidth]{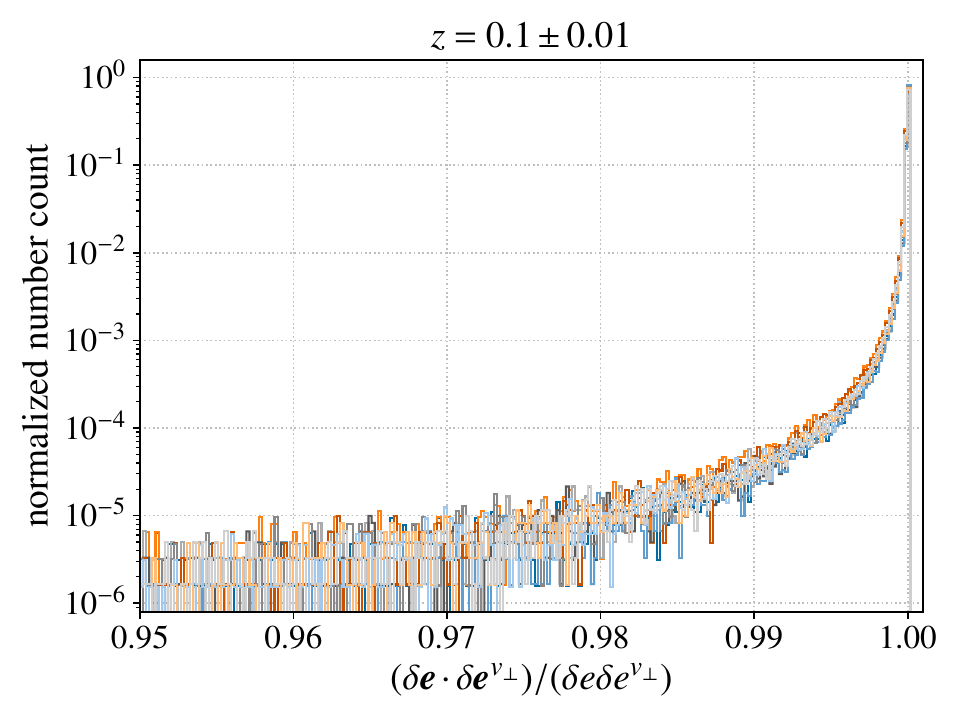}%
	\includegraphics[width=0.33\linewidth]{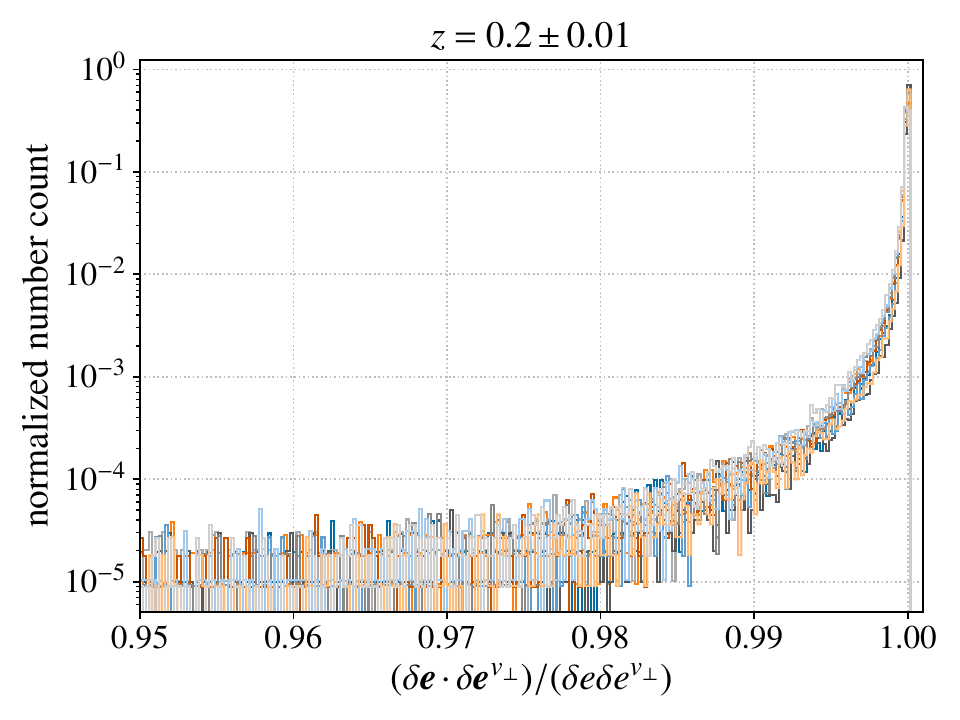}%
	\includegraphics[width=0.33\linewidth]{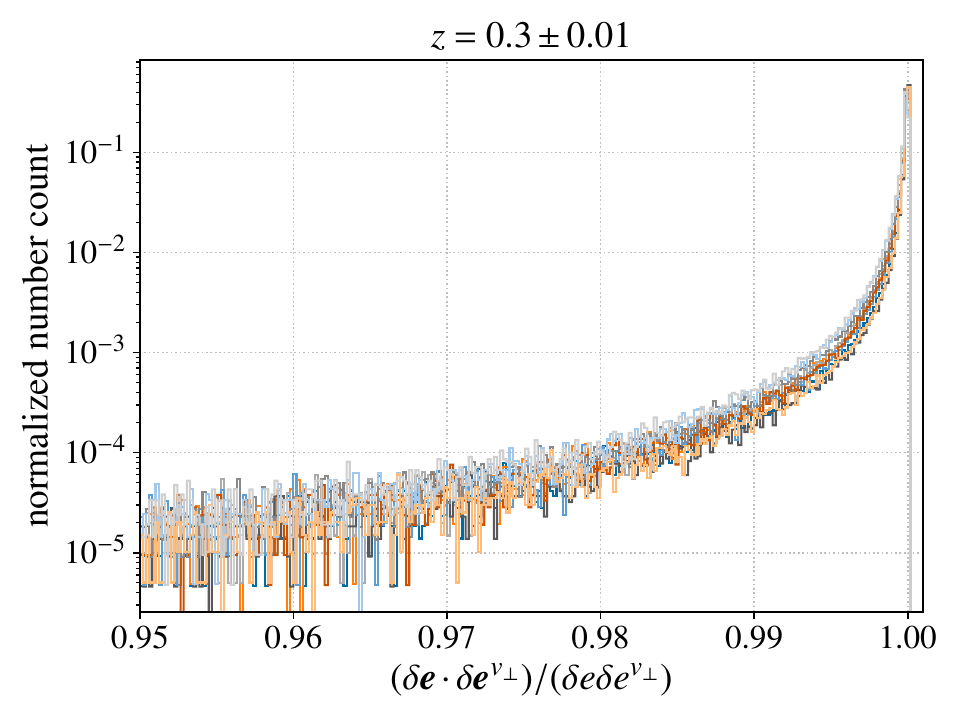}
	\caption{Normalized histograms of the relative difference between the simulation result and theory prediction according to \eqref{eq:position_drift_pertb_theory} for the position drift. The top row shows the relative difference in absolute magnitude. The bottom row shows the cosine of the angle between the two vectors. Shown for three different redshift bins $z= 0.1,0.2,0.3\pm0.01$ and ten different observers.}
    \label{fig:histograms_pertubation_theory}
\end{figure*}

In the central panel, we show the absolute value of the position drift calculated according to \eqref{eq:position_drift_pertb_theory}, i.e. the prediction from perturbation theory, but using the non-linear velocity field from the simulation as recorded on the background light cone at $z = 0$. In the right-most panel we show the difference between the two results. As seen, the theoretical prediction overall agrees well with the exact simulation results. In order to quantify how well the perturbative prediction reproduces the exact simulation result, we consider histograms of the difference between the simulation results and the prediction from perturbation theory according to \eqref{eq:position_drift_pertb_theory}. These are shown in Fig.~\ref{fig:histograms_pertubation_theory}. We quantify the difference between the two vectors via the relative difference in absolute value (upper row) and the cosine of the angle between the two vectors (bottom row). The difference is shown in three redshift bins $z= 0.1,0.2,0.3\pm0.01$ for all ten observers. 

While the position drift is well modelled by perturbation theory, there is a remaining difference of up to $\sim 5\%$ in the absolute value. Attempting to use the position drift to constrain the transverse velocity on the sky can therefore introduce an error of about $\sim 5\%$ in the velocity estimate. The magnitude of this difference seems to be largely independent of the redshift. At larger redshifts, there can be a slight offset between the theory mean and the simulation mean for some of the observers. This leads to broad flat peaks in some of the relative differences in Fig.~\ref{fig:histograms_pertubation_theory}. As can be seen from the cosine of the angle between the vectors, the two vectors are almost perfectly co-linear. 

\subsection{Power Spectra}
Using the formalism presented in Sect.~\ref{sect:3.2.1}, we now calculate E- and B-mode power spectra for the position drift. In this section, we use every tenth particle in the light cone to be able to calculate the power spectra to higher $l$. We do this using the library HEALPIX \cite{HEALPix} in the following steps: 
\begin{enumerate}
    \item Split the particles into two random subsets, each with half the particles. 
    \item Sort the particles into HEALPIX maps with $N_\textrm{side}=256$, recording the average position drift in cells with multiple particles. Create two maps each for $\delta e^\theta$ and $\delta e^\varphi$ from the two random subsets.
    \item Calculate the $\epsilon_{lm}$ and $\beta_{lm}$ from the maps for the two subsets using the HEALPIX function \texttt{map2alm\_spin}\footnote{\url{https://healpix.sourceforge.io/html/sub_map2alm_spin.htm}}. 
    \item Estimate the E- and B-mode $C_l$ by cross-correlating the two maps for $\epsilon_{lm}$ and the two maps for $\beta_{lm}$, respectively. 
    \item For redshifts with partial sky coverage, we pass a map with the empty regions masked, correct the amplitude of the power spectra by dividing by the sky fraction $C_l/f_\textrm{sky}$ and bin the spectra with $\Delta l \approx 1/f_\textrm{sky}$ to reduce noise. The binning is carried out using the script \texttt{bin\_llcl.py} included with the code \texttt{PolSpice}\footnote{\url{https://www2.iap.fr/users/hivon/software/PolSpice}} \cite{Chon2004,Szapudi2001}. The sky-fraction can be calculated from the opening half-angle $\alpha$ of the cone as\footnote{For a healpix map with finite resolution there is a small correction to this. We use the actual ratio of cells with data to total cells when calculating the power spectra in the code.} $f_\textrm{sky}=(1-\cos(\alpha))/2$. For $\alpha=25^\circ$, $f_\mathrm{sky}\approx 0.047$. 
\end{enumerate}
We split the data into two subsets and calculate the cross-correlation in order to reduce shot-noise. This method is known as jackknife resampling, see, e.g. \cite{Lepori2021}. 

\subsubsection{B-mode Purification}\label{sect:4.2.1}
\begin{figure*}
	\centering
	\includegraphics[width=0.33\textwidth]{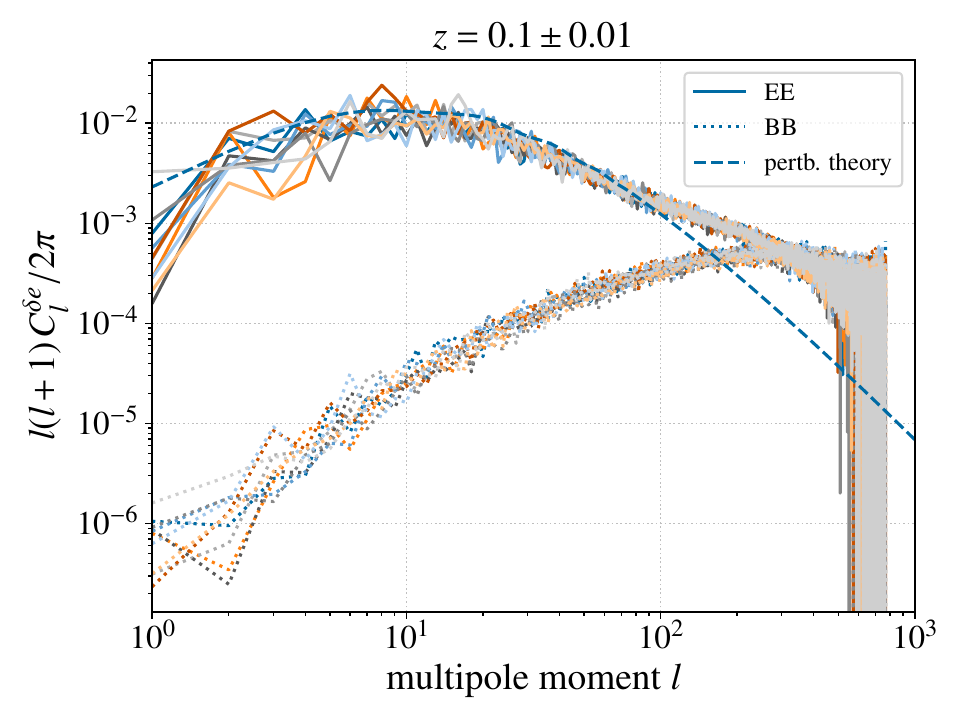}%
	\includegraphics[width=0.33\textwidth]{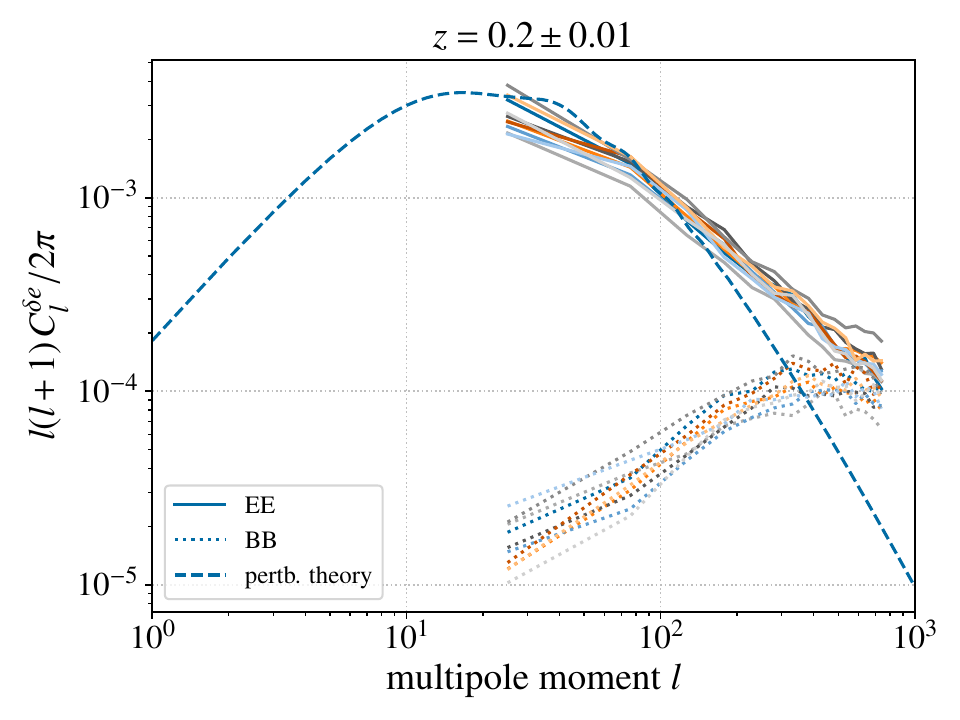}%
	\includegraphics[width=0.33\textwidth]{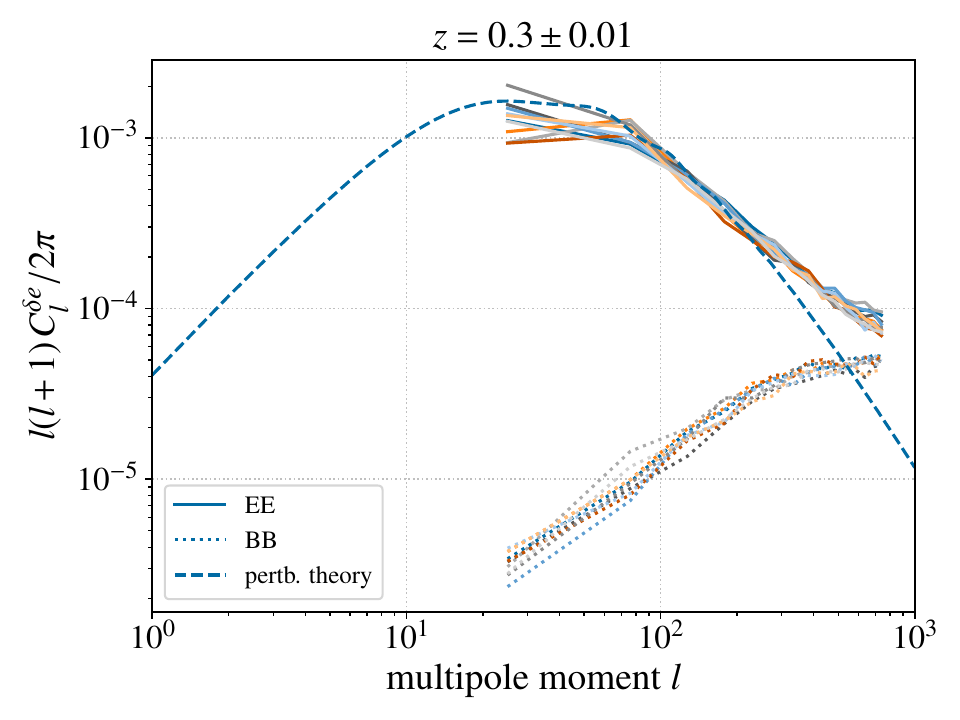}
	\caption{E- and B-mode power spectra for the position drift for three different redshifts $z=0.1,0.2,0.3\pm 0.01$ for 10 different observers. Together with the predictions for the E-mode power spectrum from linear perturbation theory according to \eqref{eq:psd_Cl}. The spectra were calculated from HEALPIX maps with side length $N_\textrm{side}=256$.}
    \label{fig:power_spectra}
\end{figure*}

While the above approach works well for the full sky map available at low redshifts, the reconstruction does not work well for the partial sky maps at higher redshifts. For the partial sky map, part of the E-mode signal leaks into the B-mode, thereby artificially increasing the B-mode amplitude for large $l$. This leakage occurs at the edge of the partial sky maps. To amend this we additionally undertake the following steps for the partial sky maps
\begin{enumerate}
    \setcounter{enumi}{5}
    \item Use the $\epsilon_{lm}$, $\beta_{lm}$ to reconstruct sky maps for the E- and B-mode. 
    \item Mask the maps to exclude the edges where the leakage occurred. We masked the original circular sky region of radius $25^\circ$ to disks with radius $22^\circ,20^\circ,18^\circ,16^\circ,14^\circ$ and found that the spectra have converged at $16^\circ$. 
    \item Use the new purified maps to recalculate the E- and B-mode $C_l$.
\end{enumerate}
The resulting power spectra can be seen in Fig.~\ref{fig:power_spectra}, together with the prediction for the E-mode $C_l$ from perturbation theory according to \eqref{eq:psd_Cl}. As seen, the E-mode is very well predicted by perturbation theory on large scales (small $l$). The B-mode is orders of magnitude smaller than the E-mode power spectrum as expected, since it vanishes identically at linear order. On small scales (large $l$), where non-linear structures build up, the E-mode mode rises above the linear theory and the B-mode is of the same order of magnitude as the E-mode. This is the expected result as vorticity builds up when non-linear structures form. 

\begin{figure*}
	\centering
	\includegraphics[width=0.33\textwidth]{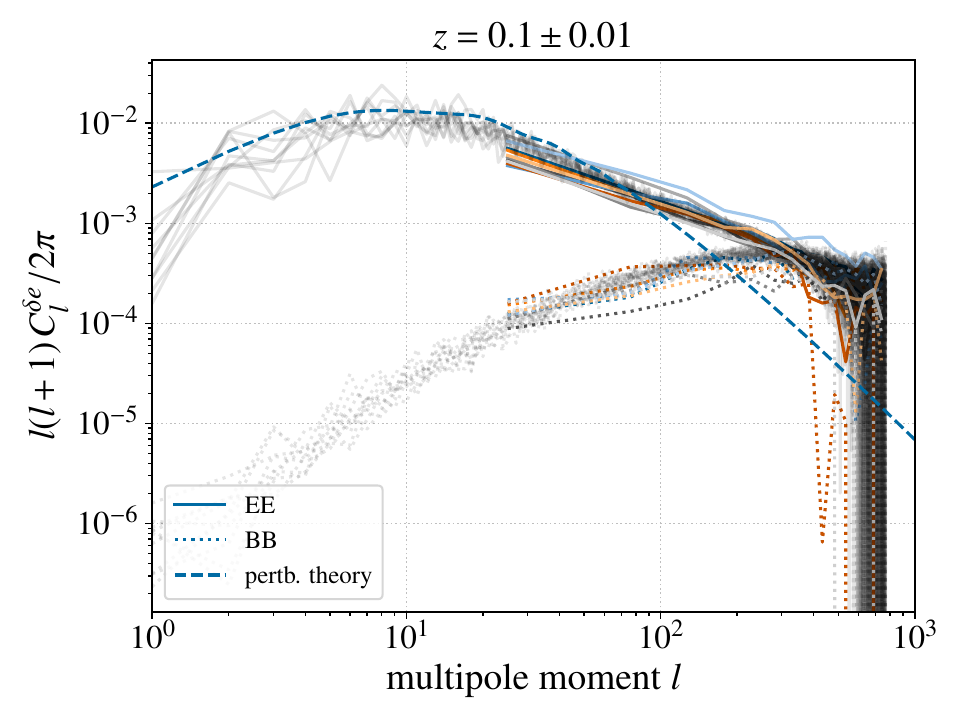}%
	\includegraphics[width=0.33\textwidth]{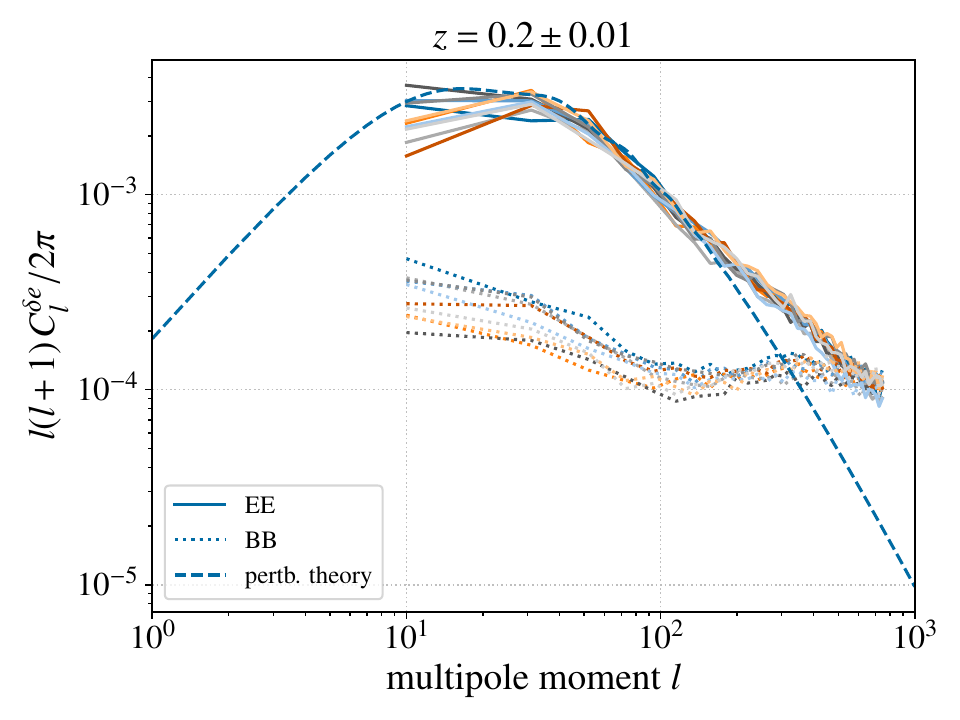}%
	\includegraphics[width=0.33\textwidth]{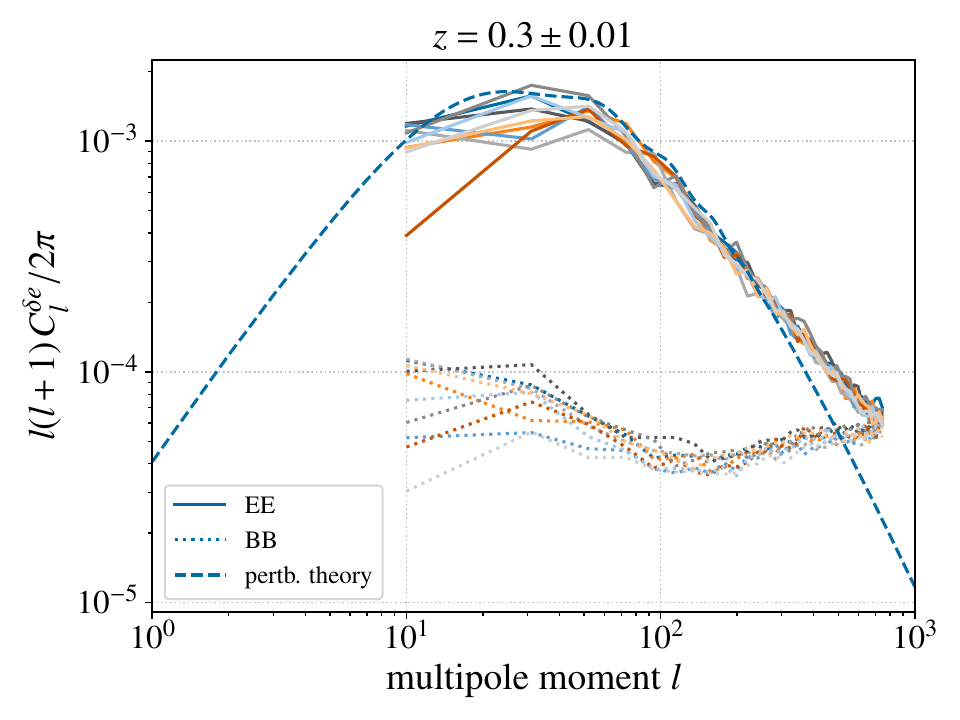}
	\caption{\textbf{Left panel:} E- and B-mode power spectra for the position drift for redshift $z=0.1\pm 0.01$ for 10 different observers. Once calculated from the full sky map (light grey curves) and once calculated from a partial masked sky map using our purification procedure described in Sect.~\ref{sect:4.2.1} (coloured curves). \textbf{Right panel:} E- and B-mode power spectra for the position drift for redshifts $z=0.2,0.3\pm 0.01$ for 10 different observers without B-mode purification. \textbf{All:} Together with the predictions for the E-mode power spectrum from linear perturbation theory according to \eqref{eq:psd_Cl}. The spectra were calculated from HEALPIX maps with side length $N_\textrm{side}=256$.}
    \label{fig:power_spectra_puryfication_verification}
\end{figure*}

\begin{figure*}
	\centering
	\includegraphics[width=0.4\textwidth]{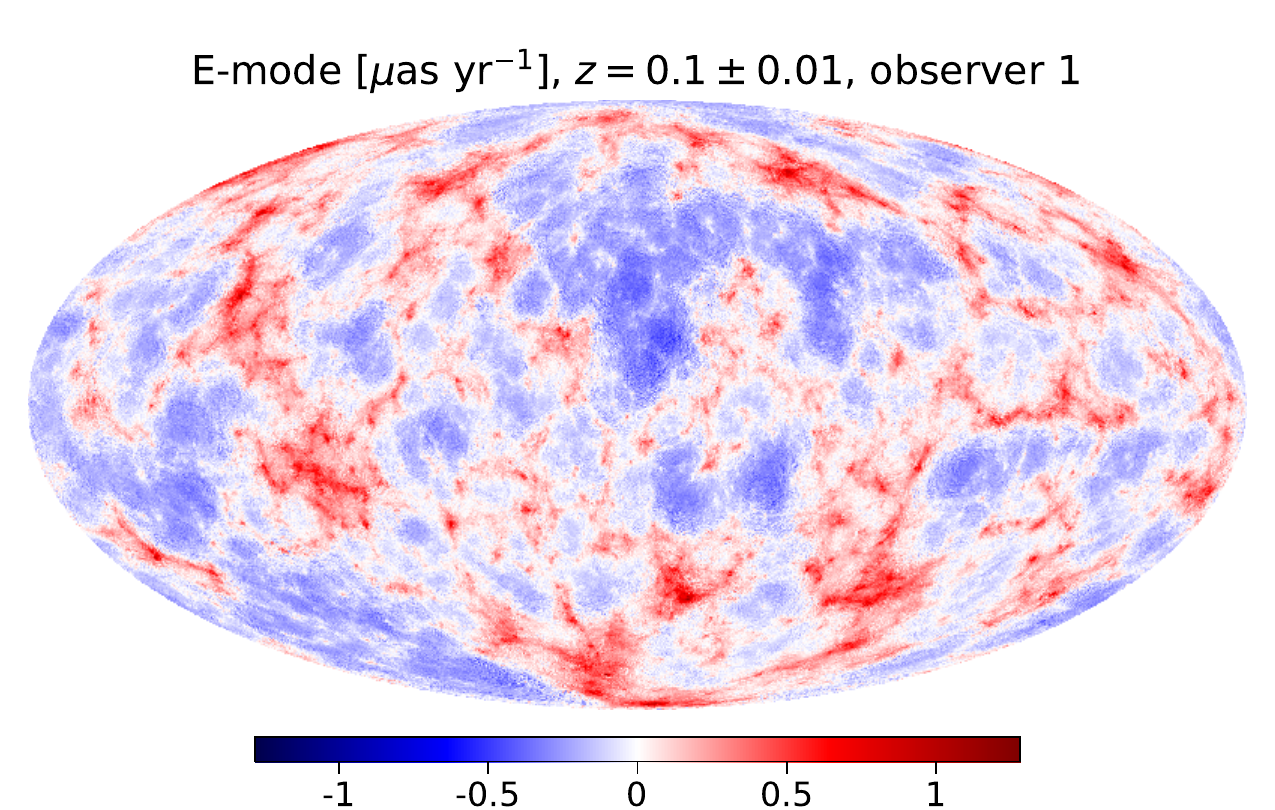}%
	\includegraphics[width=0.24\textwidth]{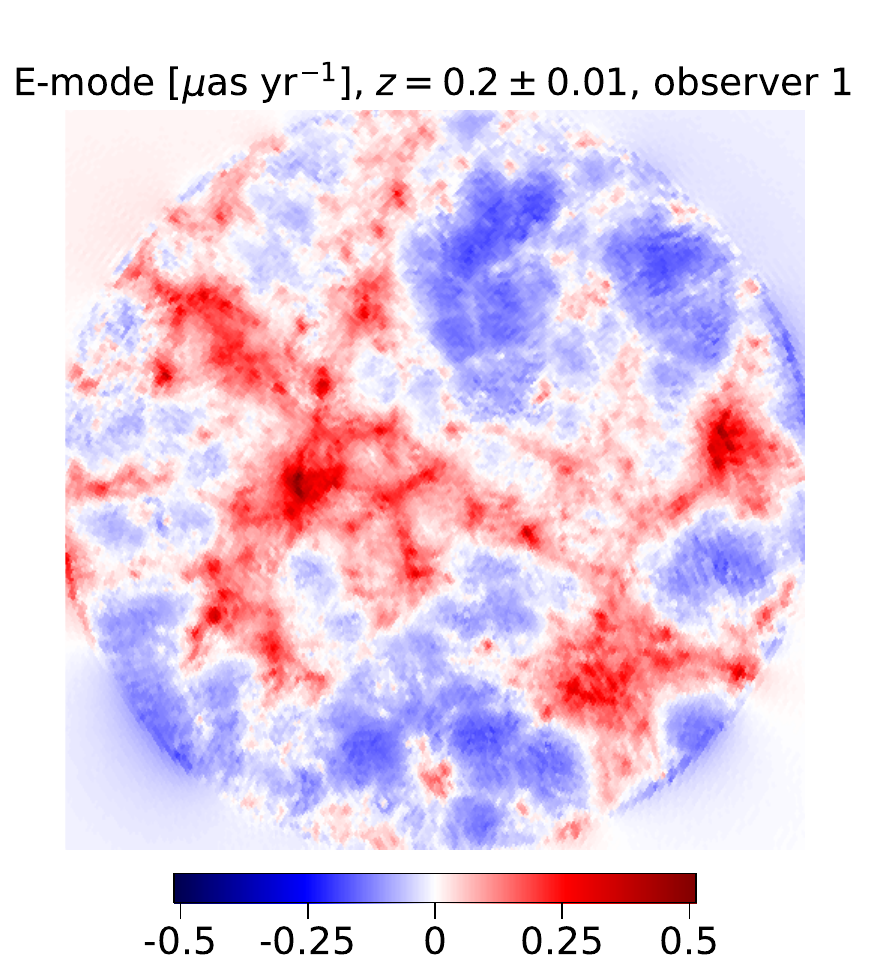}%
	\includegraphics[width=0.24\textwidth]{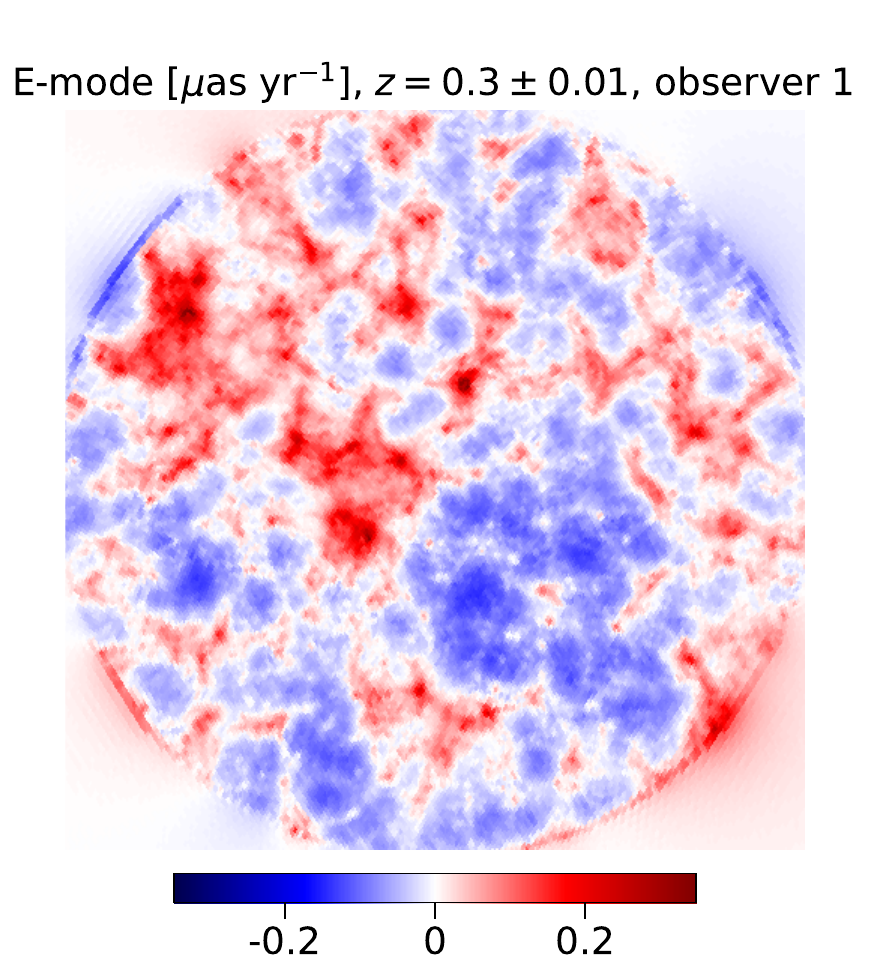}
	\includegraphics[width=0.4\textwidth]{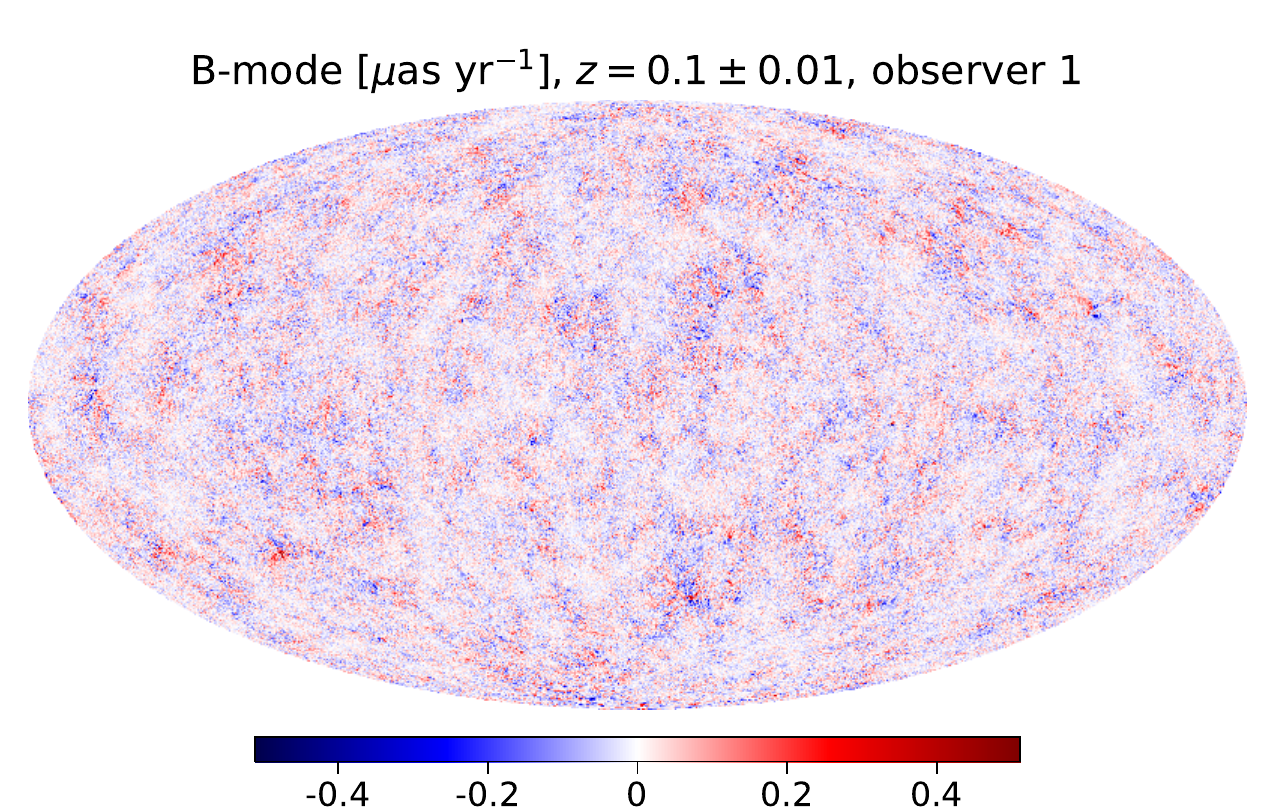}%
	\includegraphics[width=0.24\textwidth]{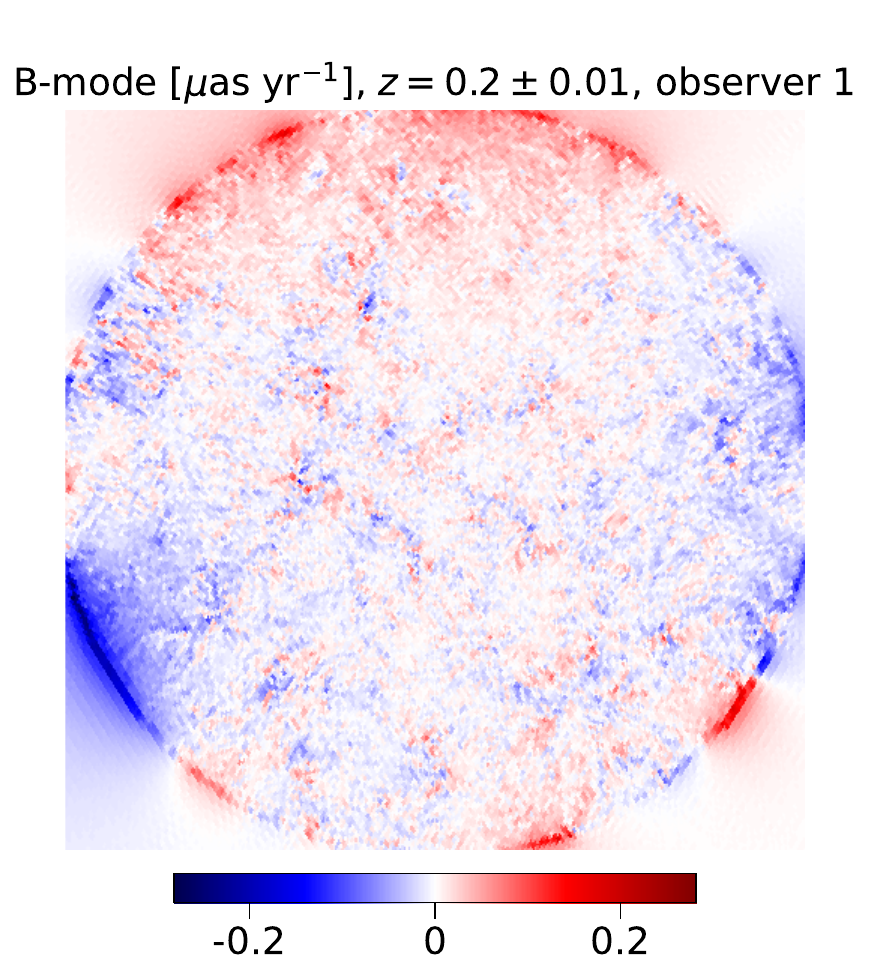}%
	\includegraphics[width=0.24\textwidth]{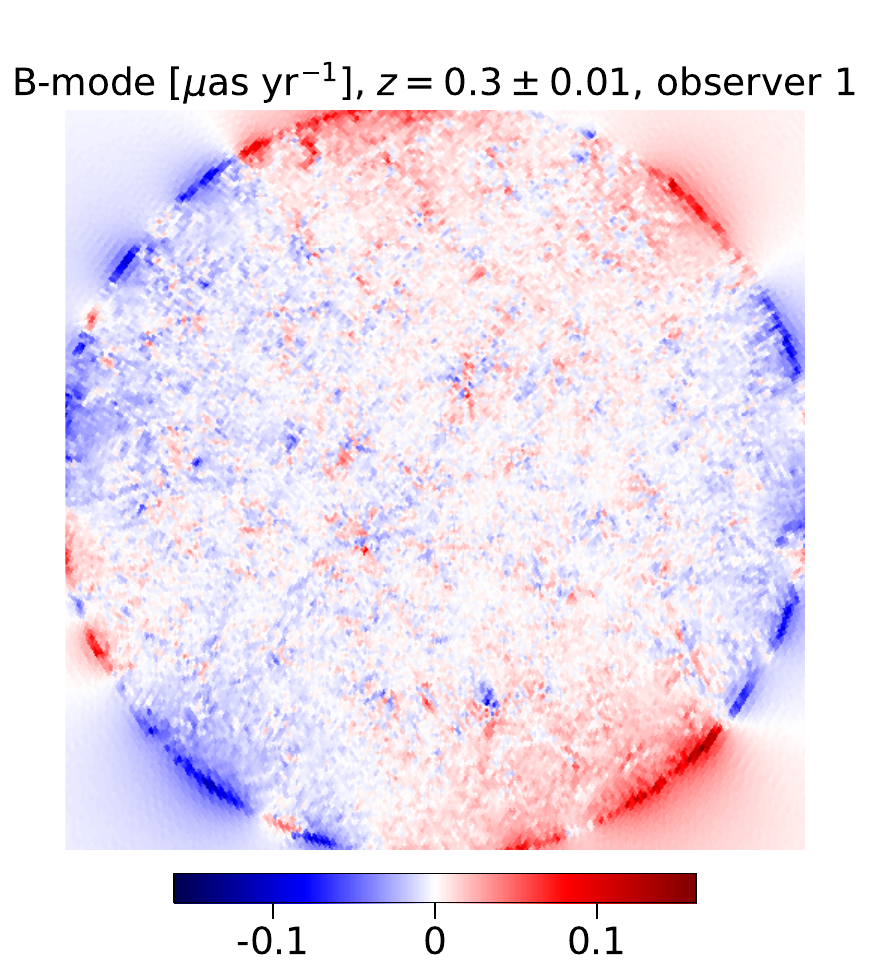}
	\caption{HEALPix skymaps of the position drift E- and B-mode, for three different redshift bins $z=0.1,0.2,0.3\pm0.01$. The map for $z\sim 0.1$ shows the full sky, while the maps for $z\sim 0.2,0.3$ show partial circular sections of the sky from a pencil beam light cone with half-opening angle 25°. Both have resolution $N_\mathrm{side}=256$.}
	\label{fig:skymaps_EB}
\end{figure*}

In order to confirm that our purification procedure works, we apply a circular mask, of the same size as our pencil beam maps, to our full sky map and calculate the power spectra from this partial map, using the above purification procedure. The resulting spectra are then compared to those calculated with the full-sky map (for those $l$ that can be recovered with the partial test map). The result is shown in the left panel of Fig.~\ref{fig:power_spectra_puryfication_verification}. As seen, the power spectra reconstructed from the partial sky (coloured curves) lie nicely on top of those from the full-sky map (light grey curves). In the same figure, we also show the power spectra for $z=0.2,0.3\pm 0.01$ calculated without the purification procedure (central and right panel). The increased B-mode signal for large $l$ caused by the E-mode leakage is clearly visible.
\\ \\ 
We show the sky maps of the E- and B-mode reconstructed from the $\epsilon_{lm}$ and $\beta_{lm}$ in Fig.~\ref{fig:skymaps_EB} for one observer. The E-mode dominates on large scales and one can clearly see the expansion of voids (blue) and contraction around structures (red). The B-mode is, in general, much smaller and only becomes visible on very small scales, as expected. In the partial B-mode sky maps for $z=0.2,0.3\pm0.01$, one can clearly see the leakage from the E-mode at the edges. 

\subsubsection{Relation to Traditional Velocity Power Spectra}
In cosmology, velocity power spectra are often studied in terms of the divergence $\theta = \nabla_{\bold r} \cdot \bold v$ and vorticity $\bm\omega = \nabla_{\bold r}\times \bold v$ with the power spectra usually defined via
\begin{align}
    \langle \theta(\bold k)\theta^*(\bold k')\rangle &= (2\pi)^3 \delta_D(\bold k-\bold k')P_{\theta\theta}(k)\;, \\
    \langle \omega^i(\bold k)\omega^{*j}(\bold k')\rangle &= (2\pi)^3\delta_D(\bold k-\bold k')\frac{1}{2}\left(\delta^{ij}-\frac{k^ik^j}{k^2}\right)P_{\omega\omega}(k)\;, 
\end{align}
see, e.g. \cite{Jelic-Cizmek2018}. The divergence $\theta$ and the velocity potential $v$ we defined earlier are related by $\theta = k^2 v$ and, therefore, we simply have $P_{\theta\theta} = k^4P_{vv}$. The vorticity power spectrum and the poloidal and toroidal spectra we defined are related in a nontrivial manner. Since we are considering observations on the past light cone, there is no clear split between the contributions from $v$ and $\bold v_R$. Instead, we can split the signal into an E- and B-mode as we introduced earlier. The E-mode signal has contributions from both $v$ and a part of the rotational part of the velocity field $\bold v_R$, namely the poloidal part $v_P$. In other words, through the projection of the velocity field onto the sphere, part of the rotational of the velocity field now contributes to the gradient-like E-mode. 

The divergence power spectrum typically falls below the predictions from linear perturbation theory on non-linear scales and is overtaken by the vorticity power spectrum, see e.g. the simulation results in \cite{Jelic-Cizmek2018,Hahn2015}. This can be explained by angular momentum conservation preventing further infall of matter and thereby converting some of the power from the divergence into vorticity as pointed out in \cite{Jelic-Cizmek2018}. Observing on the past light cone, this effect is no longer visible in the power spectra and the E-mode does not fall below the linear prediction, presumable because of the contribution of the poloidal part of the velocity field. Instead, the E- and B-modes end up with roughly the same amplitude on the smallest scales that we can resolve.

\subsubsection{Resolution Study}
\begin{figure}
    \centering
    \includegraphics[width=0.85\linewidth]{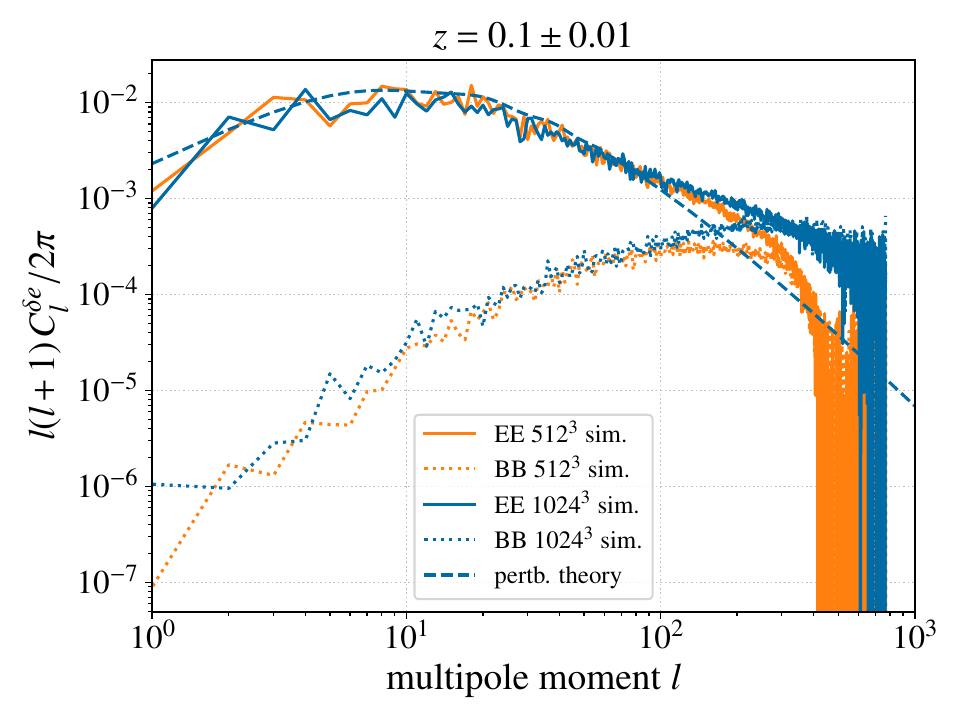}
    \caption{Angular power spectra for the position drift E- and B-mode at redshift $z=0.1\pm0.01$ for two different simulations, as well as the prediction for the E-mode from linear perturbation theory. Our main simulation (blue) and a simulation with half the resolution (orange). }
    \label{fig:power_spectra_resolution_study}
\end{figure}

Calculating velocity field power spectra from N-Body simulation is difficult and depends strongly on the resolution of the simulation. In particular on large scales the amplitude of the vorticity power spectrum is resolution dependent and converges only slowly, see e.g. \cite{Jelic-Cizmek2018,Hahn2015}. To ensure that our spectra have converged, we run a simulation with half the resolution ($512^3$ grid cells and particles) and compare the power spectra from the two simulations. We focus on one observer in the full-sky case at redshift $z=0.1\pm0.01$. In order to have sufficient particles we use all the particles in the light cone of the lower resolution simulation. The results can be seen in Fig.~\ref{fig:power_spectra_resolution_study}.  We can see that the spectra have converged on large scales (up to the expected noise). On small scales the lower resolution simulation spectra are lacking power compared to the higher resolution ones, as would be expected since the lower resolution simulation has less non-linear structure. 

\subsubsection{Dipole Amplitude}
As mentioned in the introduction, the two recent studies \cite{Makarov2025} and \cite{TsigkasKouvelis2025} found signs of redshift evolution of the position drift dipole using Gaia DR3 data. Although those results are not (yet) statistically significant, here we briefly investigate to what extent such a redshift dependence appears in our simulation.
\\ \\
For low redshifts, where we have full sky data, it is easy to calculate the dipole amplitude for the position drift. We do so by sorting the particles in a number of very low redshift bins into HEALPIX maps with side length $N_\textrm{side}=8$ and calculating the E- and B-mode $C_l$ from them. For simplicity we now again use only every one-hundredth particle in the light cones and omit the jackknife resampling. Since we are only interested in the dipole, choosing low-resolution maps, using fever particles, and not removing shot noise does not affect the results. We have mentioned before that the dipole amplitude is equal to the magnitude of the glide introduced in Sect.~\ref{sect:3.2.2} and therefore we can directly calculate it from $C_1^E$ using \eqref{eq:rot_glide_amp}. 

The results can be seen in Tab.~\ref{tab:dipole_amps} where we show the dipole amplitudes for each of the ten considered observers in different redshift intervals. Although the dipole can be quite large for very low redshifts $G\approx0.22-1.193\;\mu \textrm{as}\,\textrm{yr}^{-1}$ in the lowest redshift bin $z=0.02\pm0.01$, it quickly decays and is only $G\approx0.005-0.022\;\mu \textrm{as}\,\textrm{yr}^{-1}$ in the highest redshift bin $z=0.14\pm0.01$. Considering one very large redshift bin $z=0.07\pm0.07$, the dipole amplitude is in the interval $G \approx0.008-0.091\;\mu \textrm{as}\,\textrm{yr}^{-1}$. 

Overall, we see that the amplitude of the dipole changes by around two orders of magnitude between the considered redshift intervals, when comparing across all observers. Although we do not show the results here, we note that for some of the observers, the direction of the dipole changes significantly between the different redshift bins, while for others it remains roughly the same. 
\\ \\
These results indicate that there can be a significant observer-dependent bias in estimations of the solar system's acceleration w.r.t. the cosmic rest frame if a sample with many low-redshift sources is used. The possibility of this was already pointed out in \cite{Hall2019} and recently discussed as the ``clustering dipole'' in \cite{Rasouli2025}. However, our results also show that as soon as higher-redshift sources are included in the given redshift bin, the bias becomes quite small very quickly. We therefore do not find a significant indication that such a bias from large-scale structures can be responsible for the potential redshift evolution of the dipole in \cite{Makarov2025,TsigkasKouvelis2025}. A firm conformation of this would require considering realistic observer positions, rather than random positions as we have used here, and using a halo finder to identify realistic sources. It would furthermore require a careful investigation of the redshift distribution of sources. Indeed, when computing the dipole in the large redshift interval, we have a significant over-representation of high-redshift sources due to the geometry of the light cone. Furthermore, lower redshift sources observed with Gaia have smaller proper motion uncertainties because they are brighter and have larger proper motions. This increases the statistical weights of the low-redshift sources in analyses. This should be taken into account in a more realistic analysis.

\begin{table*}
  \centering
  \begin{tabular}{c|cccccccc}
    z & $0.02 \pm 0.01$ & $0.04 \pm 0.01$ & $0.06 \pm 0.01$ & $0.08 \pm 0.01$ & $0.10 \pm 0.01$ & $0.12 \pm 0.01$ & $0.14 \pm 0.01$ & $0.07 \pm 0.07$ \\
    \hline
    Obs.~0 & 0.341 & 0.116 & 0.088 & 0.051 & 0.031 & 0.018 & 0.012 & 0.024 \\
    Obs.~1 & 0.326 & 0.119 & 0.055 & 0.029 & 0.019 & 0.012 & 0.011 & 0.022 \\
    Obs.~2 & 0.376 & 0.166 & 0.082 & 0.037 & 0.023 & 0.011 & 0.005 & 0.028 \\
    Obs.~3 & 0.449 & 0.105 & 0.055 & 0.027 & 0.014 & 0.010 & 0.007 & 0.016 \\
    Obs.~4 & 0.766 & 0.269 & 0.122 & 0.056 & 0.026 & 0.014 & 0.010 & 0.044 \\
    Obs.~5 & 0.802 & 0.259 & 0.095 & 0.042 & 0.023 & 0.018 & 0.017 & 0.043 \\
    Obs.~6 & 0.510 & 0.211 & 0.062 & 0.044 & 0.037 & 0.023 & 0.016 & 0.036 \\
    Obs.~7 & 0.414 & 0.081 & 0.023 & 0.017 & 0.019 & 0.015 & 0.012 & 0.018 \\
    Obs.~8 & 0.220 & 0.060 & 0.041 & 0.013 & 0.016 & 0.019 & 0.016 & 0.008 \\
    Obs.~9 & 1.193 & 0.519 & 0.244 & 0.120 & 0.063 & 0.034 & 0.022 & 0.091 \\
  \end{tabular}
  \caption{Dipole amplitudes in $\;\mu \textrm{as}\,\textrm{yr}^{-1}$ for the position drift E-mode for 10 different observers and a number of different redshift bins. Calculated from HEALPIX skymaps with $N_\textrm{side}=8$.}
  \label{tab:dipole_amps}
\end{table*}

\section{Conclusions}\label{sect:5}
With the Gaia telescope and VLBI methods, it is now possible to constrain the real-time change in the position of cosmological objects. This is known as the position drift. The position drift has both a local contribution from our motion with respect to the cosmic rest frame and a cosmological contribution from the peculiar motion of objects in the universe. We have here studied the cosmological contribution to the position drift using the relativistic N-body code \texttt{gevolution}. We simulated a $\Lambda$CDM universe and calculated the position drift directly from the past light cone without approximations for ten different co-moving observers. This is the first fully non-linear and relativistic study of the position drift. 

At linear order in perturbation theory, neglecting metric fluctuations, the position drift is directly proportional to the peculiar transverse velocity of sources in the sky and thereby offers insights on the nature of gravity, dark matter and dark energy. For individual sources, this linear description is able to reproduce our non-linear simulation results to within about $5\%$. The position drift therefore offers one of the few methods for measuring the transverse velocity field; for a recent summary of complementary approaches and for new results highlighting additional imprints of transverse peculiar velocities in cosmological observables, see \cite{Cai2025}.

Using the spin-weighted spherical harmonics, we decomposed the signal into a gradient-like E- and a curl-like B-modes and calculated the corresponding angular power spectra. As expected, the E-mode dominates on large scales. On small, non-linear scales, the B-mode reaches similar amplitudes as the E-mode because vorticity builds up. On large scales, the E-mode power spectrum shows excellent agreement with the prediction from linear perturbation theory. On small scales, as non-linear structure builds up, the E-mode rises above the linear signal.

In the existing literature, the velocity power spectrum is usually decomposed into a divergence and vorticity spectrum, with the former falling off on non-linear scales and the later overtaking it on the smallest scales. We showed that when taking into account that observations are made on the past light cone, the two contributions get mixed, with part of the vorticity contributing to the E-mode.  

At low redshifts, we calculated the contribution from large-scale structure to the dipole of the position drift. Although there is a significant variation in amplitude and direction between observers, the contribution is too small to explain the possible evolution of the dipole recently reported in the literature \cite{Makarov2025,TsigkasKouvelis2025}. Our results thus strongly indicate that the reported dipole evolution cannot be attributed to large-scale structure in a $\Lambda$CDM universe, although a firm confirmation of this would require taking into account realistic observer and source positions and observation errors.

The position drift promises a complementary approach to measuring the cosmological large-scale peculiar velocity field, and it is now becoming possible to probe it with great precision. While it is still uncertain whether the current generation of telescopes is sufficient to detect the cosmological signal, the next generation (such as long-baseline ngVLA \cite{Darling2018b}) will clearly reach the required sensitivity. By providing the first fully relativistic and non-linear simulations of the effect, our work lays the groundwork for interpreting such observations and for quantifying the imprint of non-linear structure formation on the position drift signal.

\acknowledgments
AO would like to thank Queen Mary University of London for hospitality during the early stages of this project. SMK and AO are funded by VILLUM FONDEN, grant VIL53032. JA acknowledges financial support from the Swiss National Science Foundation and from the Dr.\ Tomalla Foundation for Gravity Research. The simulations used in this paper were run using the UCloud interactive HPC system managed by the eScience Centre at the University of Southern Denmark.
\newline\newline
{\bf Author contribution statement}
The analytical derivations were carried out by AO. AO also performed the numerical work with guidance from the other authors. All authors have contributed to the interpretation of the results and the development of the project. The writing of the manuscript was led by AO, but all authors contributed to the final manuscript.

\bibliography{main}

%\widetext
\end{document}